\documentclass[journal]{IEEEtran}

\usepackage[T1]{fontenc}
\usepackage[utf8]{inputenc}

\usepackage{amsmath}
\usepackage{amssymb}
\usepackage{amsfonts}
\usepackage{mathtools}
\usepackage{bm}            
\usepackage{amsthm}

\theoremstyle{plain}
\newtheorem{theorem}{Theorem}
\newtheorem{lemma}{Lemma}
\newtheorem{corollary}{Corollary}
\newtheorem{proposition}{Proposition}
\newtheorem{assumption}{Assumption}
\newtheorem{remark}{Remark}

\theoremstyle{definition}

\theoremstyle{remark}

\usepackage{tabularx}
\usepackage{multirow}
\usepackage{threeparttable}
\usepackage{booktabs}
\usepackage{array}
\usepackage{siunitx}

\usepackage{algorithm}
\usepackage{algpseudocode}

\usepackage{graphicx}
\usepackage{subcaption}

\usepackage{cite}
\usepackage{url}
\usepackage{hyperref}
\usepackage{xcolor}

\newcommand{\neff}{n_{\mathrm{eff}}}
\newcommand{\ngr}{n_{g}}
\newcommand{\Bcoh}{B_{\mathrm{coh}}}
\newcommand{\Bres}{B_{\mathrm{res}}}
\newcommand{\Babs}{B_{\mathrm{abs}}}
\newcommand{\Buse}{B_{\mathrm{use}}}
\newcommand{\Wsp}{W_{\mathrm{spatial}}}
\newcommand{\etawb}{\eta_{\mathrm{wb}}}
\newcommand{\sinc}{\operatorname{sinc}}
\usepackage{pifont}
\newcommand{\cmark}{\ding{51}}        
\newcommand{\pmark}{(\ding{51})}      
\newcommand{\xmark}{\ding{55}}        

\usepackage[cmintegrals]{newtxmath}

\def\BibTeX{{\rm B\kern-.05em{\sc i\kern-.025em b}\kern-.08em
    T\kern-.1667em\lower.7ex\hbox{E}\kern-.125emX}}

\makeatother

\makeatletter
\def\ps@IEEEtitlepagestyle{%
  \def\@oddfoot{\mycopyrightnotice}%
  \def\@oddhead{\hbox{}\@IEEEheaderstyle\leftmark\hfil\thepage}\relax
  \def\@evenhead{\@IEEEheaderstyle\thepage\hfil\leftmark\hbox{}}\relax
  \def\@evenfoot{}%
}
\makeatother

\def\mycopyrightnotice{%
  \begin{minipage}{\textwidth}
  \centering \scriptsize
    THIS WORK HAS BEEN SUBMITTED TO THE IEEE FOR POSSIBLE PUBLICATION. COPYRIGHT MAY BE TRANSFERRED WITHOUT PRIOR NOTICE.
  \end{minipage}
}

\makeatother

\begin{document}

\title{Wideband Terahertz Pinching-Antenna Systems (PASS): Bandwidth Limits, Ergodic Rate, and Placement Optimization}

\author{
        Almutasem Bellah Enad, \IEEEmembership{Graduate Student Member, IEEE,}
        Chintha Tellambura, \IEEEmembership{Fellow, IEEE,}\\%
        and Zhiguo Ding, \IEEEmembership{Fellow, IEEE.}%
\thanks{ 
Almutasem~Bellah~Enad and Chintha Tellambura are with the Department of Electrical and Computer Engineering, University of Alberta, Edmonton, AB, T6G 2R3, Canada. (e-mail: almutasem.enad@ualberta.ca, ct4@ualberta.ca). Zhiguo Ding is with the School of Electrical and Electronic Engineering (EEE), Nanyang Technological University, Singapore 639798 (e-mail: zhiguo.ding@ntu.edu.sg).}}

\maketitle

\begin{abstract}
Terahertz (THz) bands offer the vast bandwidth that sixth-generation (6G) networks demand, but their severe path loss makes bringing the radiator close to the user decisive. Pinching-antenna systems (PASS) do exactly this, turning path loss into a programmable parameter by sliding radiating points along a dielectric waveguide, which makes THz their native regime. That regime, however, is inherently wideband and the waveguide is dispersive, whereas every existing PASS analysis is single-frequency and assumes a non-dispersive guide. This paper develops the first wideband, dispersion-aware theory of PASS and uncovers a limitation intrinsic to the pinching mechanism. Because co-phasing by physical position equalizes path length only modulo a wavelength, a pinching array is inherently beam-squint limited, and we show its frequency response is a Dirichlet kernel with a closed-form coherence bandwidth and an irreducible dispersion floor that even true-time-delay (TTD) correction cannot remove. A single unifying law governs the behavior, in which array gain trades against both coherence bandwidth and spatial robustness through one invariant that TTD correction breaks, reaching the floor with far fewer delay elements than a conventional array. We further characterize the wideband ergodic rate under random user locations, show that molecular absorption acts as a deterministic taper that multiplies the squint, and derive a simple rule for the rate-optimal number of pinches. Finally, by exploiting the integer structure of co-phasing, we recast the non-convex placement problem as a mixed-integer program that can be solved exactly in polynomial time. Together, these results move PASS from the narrowband, dispersion-blind setting to the wideband, dispersive regime that THz operation actually requires.
\end{abstract}


\section{Introduction}
\IEEEPARstart{T}{he} 
Sixth-generation (6G) wireless networks envision
terabit-per-second connectivity and high-resolution
sensing, motivating the use of the terahertz (THz)
band ($0.1$--$10$~THz) and its vast available
bandwidth~\cite{sarieddeen2021overview}.
However, THz propagation faces two fundamental
limitations: severe attenuation and blockage
sensitivity. Free-space spreading and molecular
absorption substantially constrain communication
range~\cite{sarieddeen2021overview,tarboush2021teramimo,11072420},
while quasi-optical propagation and weak diffraction
produce deep shadows, with human-body blockage
causing tens of decibels of additional
loss~\cite{sarieddeen2021overview}.
These challenges motivate short, reconfigurable
line-of-sight (LOS) links that mitigate propagation
loss without requiring prohibitively large
antenna arrays.

Existing technologies address these limitations
only partially. Ultra-massive MIMO and hybrid
beamforming increase array gain
~\cite{sarieddeen2021overview,enadMImo,dai2021delay},
while holographic surfaces improve aperture
utilization~\cite{huang2020holographic}.
However, these co-located architectures primarily
concentrate energy along existing propagation
paths and cannot independently bypass LOS
blockages. Reconfigurable intelligent surfaces
(RISs) can establish alternative paths around
obstructions, but incur cascaded propagation
losses~\cite{direnzo2020smart}.
Fluid and movable antennas offer spatial
adaptation by repositioning radiating elements
~\cite{ma2024mimo,dkhan2026physicallyconsistentchannelmodeling}, yet their typically
wavelength-scale movement primarily exploits
small-scale channel variations rather than
substantially reducing propagation distance.
Consequently, these approaches do not provide
the meter-scale flexibility needed to
dynamically shorten the serving link and
mitigate large-scale THz attenuation.

In pinching-antenna systems (PASS), signals propagate
along a low-loss dielectric waveguide and are radiated
toward users through reconfigurable pinching elements
placed at selected positions. This enables on-demand
LOS links and makes large-scale path
loss controllable through antenna placement
~\cite{liu2025pass,ding2024flexible}.
By repositioning radiating points over meter-scale
distances, PASS directly address the severe
distance-dependent attenuation of THz channels.
Crucially, this mobility provides \emph{placement
freedom} rather than a correspondingly large
coherent aperture, since the radiating cluster
remains centimeter-scale. However, most existing
PASS analyse assume narrowband operation over
a non-dispersive waveguide, an assumption that
becomes increasingly restrictive for wideband
THz transmission.

\subsection{Related Work}
Table~\ref{tab:literature} compares this work with
representative PASS and wideband-array studies,
highlighting a clear research gap between two
largely separate lines of investigation that
the proposed framework unifies.

 The first investigates PASS primarily under
narrowband propagation models. Early works
established closed-form rate expressions and
multiuser beamforming through antenna placement
~\cite{ding2024flexible}, uplink placement and
resource allocation~\cite{tegos2025minimum},
NOMA-assisted activation~\cite{wang2024noma},
array-gain and coupling limits
~\cite{ouyang2025array}, and architectural
principles~\cite{liu2025pass}.
Subsequent studies extended PASS to MIMO
beamforming~\cite{bereyhi2025mimopass},
capacity analysis~\cite{ouyang2025capacity},
and robust degree-of-freedom optimization
~\cite{dofrobust2026}.
Moreover,
\cite{tyrovolas2026performance} incorporated
waveguide attenuation and random user
locations, but retained a single-frequency,
single-pinching-antenna model.
These studies do not explicitly characterize
frequency-dependent guided-wave phase
propagation or its implications for
wideband beamforming.

The second direction addresses wideband
beam squint in conventional phased and
hybrid arrays. Frequency-independent phase
shifters produce frequency-dependent beam
directions, degrading coherent array gain
over large bandwidths
~\cite{dai2021delay,tarboush2021teramimo}.
True-time-delay (TTD) and delay-phase
precoding mitigate this effect by providing
frequency-dependent phase compensation
~\cite{dai2021delay}, with recent extensions
to flexible-antenna architectures
~\cite{sixdma2026}.
However, these approaches do not account
for the frequency-dependent phase accumulated
along a dielectric waveguide, which governs
coherent radiation in PASS.
Consequently, the interaction between
waveguide dispersion, wideband beam squint,
and pinching-antenna placement remains
largely unexplored. In particular, it is
unclear how dispersion affects the beam
direction and coherent array gain across
frequency, limits the usable bandwidth,
and influences optimal pinching positions.
This paper addresses these questions
through a unified wideband PASS framework
that explicitly incorporates guided-wave
dispersion and attenuation, thereby
bridging the two research directions
summarized in Table~\ref{tab:literature}.

\begin{table}[t]
\centering
\caption{Positioning of this work against representative wideband-array and
pinching-antenna studies.}
\label{tab:literature}
\renewcommand{\arraystretch}{1.3}
\footnotesize
\setlength{\tabcolsep}{3.0pt}
\begin{tabular}{|l|c|c|c|c|c|c|c|}
\hline
\textbf{Criterion} &
\rotatebox{0}{\cite{dai2021delay}} &
\rotatebox{0}{\cite{sixdma2026}} &
\rotatebox{0}{\cite{ding2024flexible}} &
\rotatebox{0}{\cite{ouyang2025array}} &
\rotatebox{0}{\cite{bereyhi2025mimopass}} &
\rotatebox{0}{\cite{tyrovolas2026performance}} &
\rotatebox{0}{\textbf{This}} \\
\hline\hline
Pinching antenna (PASS)      & \xmark & \xmark & \cmark & \cmark & \cmark & \cmark & \cmark \\
\hline
Wideband multicarrier        & \cmark & \cmark & \xmark & \xmark & \xmark & \xmark & \cmark \\
\hline
Waveguide dispersion         & \xmark & \xmark & \xmark & \xmark & \xmark & \xmark & \cmark \\
\hline
Beam squint / coherence BW   & \cmark & \cmark & \xmark & \xmark & \xmark & \xmark & \cmark \\
\hline
TTD squint mitigation        & \cmark & \pmark & \xmark & \xmark & \xmark & \xmark & \cmark \\
\hline
Molecular absorption         & \xmark & \xmark & \xmark & \xmark & \xmark & \xmark & \cmark \\
\hline
Random-user ergodic rate     & \xmark & \xmark & \xmark & \xmark & \xmark & \cmark & \cmark \\
\hline
Optimal placement            & \xmark & \pmark & \pmark & \pmark & \pmark & \xmark & \cmark \\
\hline
\end{tabular}
{\footnotesize \cmark~fully\quad \pmark~partially\quad \xmark~not addressed.}
\end{table}

\subsection{Contributions}
\noindent We develop the first wideband, dispersion-aware theory of PASS. Our contributions are as follows.
\begin{enumerate}

\item \textbf{Wideband dispersive channel model:} We derive a PASS channel that
jointly captures the $1/f^2$ free-space roll-off, frequency-dependent waveguide
attenuation, and waveguide dispersion, reducing the dispersion to a few physical
constants through a group-delay expansion. This is the
first PASS model valid across a wideband.

\item \textbf{Inherent beam squint and coherence bandwidth:} We prove that
single-waveguide co-phasing equalizes path length only modulo a wavelength, so a
pinching array is inherently squint-limited. The array response is a Dirichlet
kernel in frequency, which yields a closed-form coherence bandwidth and an $N$-invariant gain-bandwidth product.

\item \textbf{Irreducible dispersion floor:} We isolate the
group-velocity-dispersion (GVD) term as the single component that TTD
correction cannot remove, and derive the resulting bandwidth floor that bounds any corrected pinching array.

\item \textbf{Wideband rate, absorption, and a design rule for $\textbf{N}$:} We characterize
the ergodic rate under random user location and prove that the user-position
statistics and the wideband squint decouple, the squint entering only through a
deterministic efficiency scalar. We further show molecular absorption multiplies
this scalar as a deterministic taper with a closed-form absorption-limited
bandwidth, and we prove that the band-integrated rate is maximized when the
coherence bandwidth meets the signal bandwidth.

\item \textbf{Delay-phase mitigation:} We propose a delay-phase architecture of
coarse feed TTD plus pinch positioning that attains the dispersion floor with only
$\mathcal{O}(\sqrt{N})$ delay elements rather than one per pinch.

\item \textbf{Optimal and robust placement:} Exploiting the integer structure of
path-modulo-wavelength co-phasing, we reformulate the non-convex wideband
placement problem \emph{exactly} as a mixed-integer program, prove that
minimal-step consecutive orders are optimal under an explicit spacing-feasibility
condition so the problem is polynomially solvable, and establish a spatial
position-coherence region whose gain-robustness product mirrors the
gain-bandwidth law.
\end{enumerate}


\section{System Model and Problem Formulation}\label{sec:model}
\begin{figure}[t]
\centering
\includegraphics[width=\columnwidth]{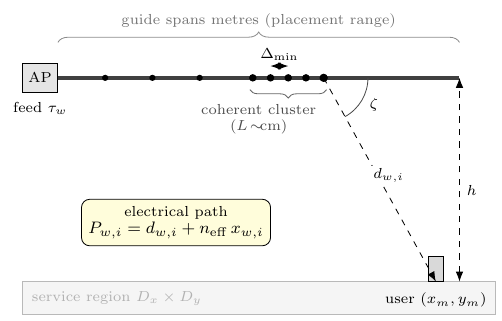}
\caption{Wideband dispersive PASS: the coherent cluster is centimeter-scale on a
meters-long guide, giving placement freedom rather than aperture. Each pinch combines a
hop $d_{w,i}$ and in-guide run $n_{\mathrm{eff}}x_{w,i}$ at angle $\zeta$.}
\label{fig:sysmodel}
\end{figure}

We introduce in this section the unified system model used throughout the paper,
summarized in Fig.~\ref{fig:sysmodel}, in which a pinch cluster far smaller than the
waveguide is positioned near the user, the feature that distinguishes PASS from a
fixed aperture. 
\subsection{Pinching-Antenna Array and Wideband Signal Model}\label{subsec:geom}
We consider a downlink in which an access point serves a single-antenna user through a
pinching-antenna array built on dielectric waveguides
\cite{liu2025pass,ding2024flexible}. The array comprises $W\ge1$
parallel dielectric waveguides indexed by $w\in\{1,\dots,W\}$. Waveguide $w$ is aligned
with the $x$ axis at height $h$ and is fed at the point $\bm{\psi}_0=(0,0,h)\in\mathbb{R}^3$
through a TTD element of delay $\tau_w\in\mathbb{R}^{+}$. It carries $N_w$
radiating points, the pinches, located at $\bm{\psi}_{w,i}=(x_{w,i},0,h)\in\mathbb{R}^3$
for $i\in\{1,\dots,N_w\}$, with axial positions ordered as
$0\le x_{w,1}<\cdots<x_{w,N_w}\le D_x$. The total number of pinches is
$N=\sum_{w=1}^{W}N_w$, and a minimum spacing $\Delta_{\min}$ separates any two pinches so
that electromagnetic coupling between radiating sites is negligible \cite{ouyang2025array}.
Unlike fluid or movable antennas, whose displacement spans only a few wavelengths and
therefore reshapes small-scale fading \cite{ma2024mimo},
a pinch can be applied anywhere along a waveguide that may extend over meters, so that the
radiating site can be brought close to the user and the large-scale pathloss becomes a
controllable quantity \cite{liu2025pass,tyrovolas2026performance}.

The single-antenna user lies in the plane $z=0$ at $\bm{\psi}_m=(x_m,y_m,0)\in\mathbb{R}^3$,
within a rectangular service region of size $D_x\times D_y$, so that $x_m\in[0,D_x]$ and
$y_m\in[-D_y/2,D_y/2]$. The free-space distance from pinch $(w,i)$ to the user is
\begin{equation}
d_{w,i}=\|\bm{\psi}_m-\bm{\psi}_{w,i}\|=\sqrt{(x_m-x_{w,i})^2+y_m^2+h^2}\in\mathbb{R}^{+}.
\label{eq:dist}
\end{equation}
In the single-waveguide case $W=1$ we drop the index $w$ and write $x_i$, $d_i$, and
$N_1=N$. This case underlies Sections~\ref{sec:squint}--\ref{sec:rate}, whereas the
multi-waveguide case supports the designs of Sections~\ref{sec:fix}--\ref{sec:opt}.

We study a wideband system of total bandwidth $B$ centered at the carrier $f_c$,
consistent with the ultra-broadband operation envisioned for THz links
\cite{rappaport2019wireless,tarboush2021teramimo,11362947}. Orthogonal frequency-division
multiplexing with $K$ subcarriers is assumed, the $k$-th subcarrier occupying frequency
\begin{equation}
f_k=f_c+\Big(k-\tfrac{K}{2}\Big)\frac{B}{K},\qquad k\in\{0,\dots,K-1\},
\label{eq:fk}
\end{equation}
and a generic in-band frequency is written $f=f_c+\Delta f$ with $|\Delta f|\le B/2$. We
denote by $c$ the speed of light and by $\lambda_c=c/f_c$ the carrier wavelength.

\subsection{Wideband Dispersive Channel Model}\label{subsec:channel}
Three physically distinct mechanisms make the link frequency selective. The narrowband
pinching-antenna literature evaluates all three at the single frequency $f_c$
\cite{ding2024flexible,ouyang2025capacity,bereyhi2025mimopass}, an approximation that is
untenable across a THz band. We retain their frequency dependence.
\subsubsection{Free-Space Spreading}
The first mechanism is free-space spreading, whose reference gain at unit distance is
\begin{equation}
\eta(f)=\frac{c^2}{16\pi^2 f^2},
\label{eq:eta}
\end{equation}
decaying as $1/f^2$ and representing the growth of free-space loss with frequency
\cite{tyrovolas2026performance}. 
\subsubsection{In-Waveguide Attenuation} 
The second mechanism is in-waveguide attenuation, modeled
by the frequency-dependent power-attenuation coefficient
\begin{equation}
\alpha(f)\approx\alpha_0\,(f/f_c)^{s},
\label{eq:alpha}
\end{equation}
where $\alpha_0\ge0$ is the attenuation at $f_c$ and $s\ge0$ captures the rise of dielectric
loss with frequency \cite{yeh2008essence,balanis2012advanced}. 
\subsubsection{Guided Propagation and Dispersion}
The third mechanism is the guided phase accumulated along the waveguide,
described by the propagation constant
\begin{equation}
\beta(f)=\frac{2\pi f\,n_{\mathrm{eff}}(f)}{c},
\label{eq:beta}
\end{equation}
which sets the phase per unit length of the guided wave and whose dependence on
frequency through the effective index $n_{\mathrm{eff}}(f)$ constitutes waveguide
dispersion \cite{yeh2008essence}.

Collecting free-space spreading, in-waveguide attenuation, the free-space propagation phase,
and the guided phase, the channel contributed by pinch $(w,i)$ at frequency $f$ is
\begin{equation}
H_{w,i}(f)=\underbrace{\frac{\sqrt{\eta(f)}}{d_{w,i}}\,e^{-\alpha(f)x_{w,i}/2}}
_{\triangleq\,A_{w,i}(f)}\,e^{-j\phi_{w,i}(f)},
\label{eq:Hwi}
\end{equation}
where $A_{w,i}(f)\in\mathbb{R}^{+}$ is the amplitude and the total phase gathers the
free-space hop and the guided run,
\begin{equation}
\phi_{w,i}(f)=\frac{2\pi f}{c}\,d_{w,i}+\beta(f)\,x_{w,i}
=\frac{2\pi f}{c}\,P_{w,i}(f).
\label{eq:phi}
\end{equation}
Here
\begin{equation}
P_{w,i}(f)=d_{w,i}+n_{\mathrm{eff}}(f)\,x_{w,i}
\label{eq:elecpath}
\end{equation}
is the electrical path of the pinch, equal to the physical free-space hop plus the
electrically scaled in-waveguide run. The combined channel at the user, including the feed
delays, is
\begin{equation}
H(f)=\sum_{w=1}^{W}e^{-j2\pi f\tau_w}\sum_{i=1}^{N_w}A_{w,i}(f)\,e^{-j\phi_{w,i}(f)}
\in\mathbb{C}.
\label{eq:Hcomb}
\end{equation}

To separate the first-order index from genuine dispersion we expand $n_{\mathrm{eff}}(f)$
about $f_c$ following standard guided-wave practice \cite{balanis2012advanced,yeh2008essence},
\begin{equation}
n_{\mathrm{eff}}(f)\approx n_{\mathrm{eff}}+n_{\mathrm{eff}}'\,\Delta f,\qquad
n_g\triangleq n_{\mathrm{eff}}+f_c\,n_{\mathrm{eff}}',
\label{eq:ng}
\end{equation}
where $n_{\mathrm{eff}}\triangleq n_{\mathrm{eff}}(f_c)$, the prime denotes differentiation
at $f_c$, and $n_g$ is the group index. A non-dispersive guide has $n_g=n_{\mathrm{eff}}$.
The group index governs delay rather than phase, a distinction that proves central in
Section~\ref{sec:both}.

For the squint analysis we adopt the following standard approximation, which isolates the
phase behavior that drives beam squint from the slowly varying amplitude
\cite{dai2021delay}.
\begin{assumption}\label{as:amp}
For pinches clustered near the pathloss-optimal site and over the coherence bandwidth of
interest, the amplitudes in \eqref{eq:Hwi} are approximately constant across pinches and
across in-band frequencies, that is $A_{w,i}(f)\approx\bar A(f)$. Amplitude taper is treated
as a second-order refinement in Section~\ref{sec:opt}.
\end{assumption}
\noindent
Under Assumption~\ref{as:amp}, a single waveguide with $W=1$ and $\tau_1=0$ yields the
normalized array gain
\begin{equation}
G(f)\triangleq\frac{|H(f)|^2}{N^2|\bar A(f)|^2}
=\bigg|\frac{1}{N}\sum_{i=1}^{N}e^{-j\phi_i(f)}\bigg|^2\in[0,1],
\label{eq:G}
\end{equation}
which equals unity when the $N$ contributions combine coherently.

\subsection{Problem Formulation}\label{subsec:problems}
With the model above, the received signal on subcarrier $k$ is $y_k=H(f_k)\,s_k+n_k$, where
$s_k$ is the transmitted symbol and $n_k\sim\mathcal{CN}(0,\sigma^2)$ is additive noise.
Under an equal power split of the total power $P$ across the $K$ subcarriers and the $N$
pinches of a passive feed, consistent with the array-gain accounting of
\cite{tyrovolas2026performance}, the received signal-to-noise ratio (SNR) on subcarrier $k$ at the
pathloss-optimal placement, where the cluster sits at $\bar x=x_m$ so that the common
distance is $d^2=y_m^2+h^2$, is
\begin{equation}
\gamma_k=\frac{P\,N}{K\sigma^2}\,
\frac{\eta(f_k)\,e^{-\alpha(f_k)x_m}\,G(\Delta f_k)}{y_m^2+h^2}.
\label{eq:snr}
\end{equation}

The paper addresses three problems built on this model. The first is the characterization of
the usable wideband range, that is the coherence bandwidth over which $G(\Delta f)$ remains within $3$~dB of its peak. The
second is the wideband ergodic spectral efficiency
\begin{equation}
\bar R=\mathbb{E}_{x_m,y_m}\!\bigg[\frac{1}{K}\sum_{k=0}^{K-1}\log_2\big(1+\gamma_k\big)\bigg],
\label{eq:rate}
\end{equation}
averaged over a random user location. The third is the
joint optimization of pinch positions and feed delays that maximizes the band-averaged
received power
\begin{equation}
\Theta\big(\{x_{w,i}\},\{\tau_w\}\big)=\frac{1}{K}\sum_{k=0}^{K-1}\big|H(f_k)\big|^2,
\label{eq:Theta}
\end{equation}
subject to the region, spacing, and causality constraints.
 All three problems hinge on a feature absent from prior
PASS analyses: the frequency dependence of $G(\Delta f)$. A single-frequency model
sees only $f_c$, where $G(0)=1$ and the array gives its full coherent gain $N$, so
rate, bandwidth, and placement reduce to their narrowband counterparts. Across a
wide band $G(\Delta f)<1$ for $\Delta f\neq0$, and the three become three views of
one object: the first \emph{characterizes} it via the coherence bandwidth $\Bcoh$
and the post-correction dispersion floor (Section~\ref{sec:both}), the second
\emph{integrates} it into the ergodic rate \eqref{eq:rate} (Section~\ref{sec:rate}),
and the third \emph{shapes} it through placement and feed delays \eqref{eq:Theta}
(Sections~\ref{sec:fix}--\ref{sec:opt}). The parts thus form one loop around the
same gain profile, not separate studies.
\section{Coherence Bandwidth and Dispersion Floor}\label{sec:both}
\subsection{Inherent Beam Squint and Coherence Bandwidth}\label{sec:squint}
This section establishes the central limitation of a pinching array over a wideband. We
show that single-waveguide co-phasing is intrinsically frequency selective, so the array is
beam-squint limited, and we obtain its coherence bandwidth in closed form. We take $W=1$
throughout.

A pinch controls phase only through its physical position, so co-phasing equalizes path
length modulo a wavelength rather than true delay. The following result makes this precise.
\begin{lemma}\label{lem:cophase}
Coherent combining of pinches $i$ and $\ell$ at the design frequency $f_c$ requires the
electrical-path difference to be an integer multiple of the carrier wavelength,
\begin{equation}
\Delta P_{i\ell}(f_c)\triangleq P_i(f_c)-P_\ell(f_c)=m_{i\ell}\,\lambda_c,\qquad
m_{i\ell}\in\mathbb{Z}.
\label{eq:cophase}
\end{equation}
At any other frequency the corresponding phase difference is
\begin{equation}
\phi_i(f)-\phi_\ell(f)=2\pi\,m_{i\ell}\,\frac{f}{f_c},
\label{eq:phasediff}
\end{equation}
which is a multiple of $2\pi$ only at $f=f_c$ and its harmonics.
\end{lemma}
\begin{IEEEproof}
From \eqref{eq:phi}, $\phi_i(f)=(2\pi f/c)P_i(f)$. Co-phasing at $f_c$ means
$\phi_i(f_c)-\phi_\ell(f_c)\in2\pi\mathbb{Z}$, which with $f_c/c=1/\lambda_c$ gives
\eqref{eq:cophase}. For a non-dispersive guide $P_i$ does not depend on frequency, so
$\phi_i(f)-\phi_\ell(f)=(2\pi f/c)m_{i\ell}\lambda_c=2\pi m_{i\ell}f/f_c$, which lies in
$2\pi\mathbb{Z}$ only when $f/f_c$ is an integer, proving \eqref{eq:phasediff}.
\end{IEEEproof}

Lemma~\ref{lem:cophase} is the crux of the wideband behavior. Because alignment is set by
path length modulo $\lambda_c$, it holds exactly only at the frequency for which it was
designed. A TTD array equalizes physical delay and would hold at all frequencies
\cite{dai2021delay}, but a single passive waveguide offers position as its only control and
cannot. The resulting squint is quantified exactly, to first order in $\Delta f$, by the
differential group delay between propagation paths.
\begin{lemma}\label{lem:groupdelay}
Let the pinches be co-phased at $f_c$, and define the differential free-space hop
$u_i\triangleq d_i-d_1$ and the differential guide length $v_i\triangleq x_i-x_1$. To first
order in $\Delta f$, the residual phase of pinch $i$ relative to pinch $1$ is
\begin{equation}
\psi_i(\Delta f)=2\pi\,\Delta f\,\tau_i,\qquad \tau_i=\frac{u_i+n_g\,v_i}{c},
\label{eq:groupdelay}
\end{equation}
where $\tau_i$ is the differential group delay. The group index $n_g$, not the phase index
$n_{\mathrm{eff}}$, sets the squint rate.
\end{lemma}
\begin{IEEEproof}
Using \eqref{eq:phi} and \eqref{eq:ng},
\begin{align}
\phi_i(f)\!-\!\phi_1(f)
\!=\!\frac{2\pi f}{c}\!\big[u_i\!+\!n_{\mathrm{eff}}(f)v_i\!\big]\!\!=\!\frac{2\pi f}{c}\!\big[m_i\lambda_c\!+\!n_{\mathrm{eff}}'\Delta f\,v_i\big],
\label{eq:gd1}
\end{align}
where the co-phasing identity $u_i+n_{\mathrm{eff}}v_i=m_i\lambda_c$ from
Lemma~\ref{lem:cophase} was used with $m_i\triangleq m_{i1}$. Subtracting the coherent part
$2\pi m_i$ and writing $f/f_c=1+\Delta f/f_c$ gives
$\psi_i(\Delta f)=2\pi m_i\Delta f/f_c+(2\pi f/c)n_{\mathrm{eff}}'\Delta f\,v_i$. To first
order we set $f\to f_c$ in the second term and use $n_{\mathrm{eff}}'=(n_g-n_{\mathrm{eff}})/f_c$
together with $2\pi m_i\Delta f/f_c=(2\pi\Delta f/c)(u_i+n_{\mathrm{eff}}v_i)$. The two terms
combine into $(2\pi\Delta f/c)(u_i+n_g v_i)$, which is \eqref{eq:groupdelay}.
\end{IEEEproof}

We now specialize to the canonical placement, in which the $N$ pinches are co-phased with
consecutive orders so that $P_i(f_c)-P_1(f_c)=(i-1)\lambda_c$. This placement uses the
densest packing that keeps the pinches co-phased
(subject to the spacing-feasibility condition of Section~\ref{subsec:feas}; order-skipping
applies when it is violated),
and Section~\ref{sec:opt} proves it optimal for the wideband gain, so it is adopted here
without loss. The next result is the central characterization of this paper.
\begin{theorem}\label{thm:bcoh}
Under Assumption~\ref{as:amp} and the canonical placement, the normalized array gain
\eqref{eq:G} is the Dirichlet kernel in frequency
\begin{equation}
G(\Delta f)=\bigg[\frac{\sin\!\big(\pi N\chi\,\Delta f/f_c\big)}
{N\sin\!\big(\pi\chi\,\Delta f/f_c\big)}\bigg]^{2},\qquad
\chi=\frac{n_g-\cos\zeta}{n_{\mathrm{eff}}-\cos\zeta},
\label{eq:dirichlet}
\end{equation}
where $\zeta$ is the angle between the waveguide axis and the line from the cluster to the
user. Its two-sided $3$-dB coherence bandwidth is
\begin{equation}
B_{\mathrm{coh}}=\frac{2z^\star}{\pi}\,\frac{f_c}{N\chi}
\approx0.886\,\frac{f_c}{N}\,\frac{n_{\mathrm{eff}}-\cos\zeta}{n_g-\cos\zeta},
\label{eq:bcoh}
\end{equation}
where $z^\star=1.3916$ is the unique root of $\sin (z)/z=1/\sqrt2$ in $(0,\pi)$. At the
broadside operating point $\cos\zeta=0$ the factor reduces to $n_{\mathrm{eff}}/n_g$, and in
a non-dispersive guide to unity.
\end{theorem}
\begin{IEEEproof}
By Lemma~\ref{lem:groupdelay}, for a small aperture the pinch-user angle $\zeta$ is nearly
constant, so $u_i\approx-\cos\zeta\,v_i$ and $(n_{\mathrm{eff}}-\cos\zeta)v_i=m_i\lambda_c$.
Substituting into \eqref{eq:groupdelay} gives the residual phase
$\psi_i(\Delta f)=2\pi m_i(\Delta f/f_c)\chi$ with $\chi$ as stated. With $m_i=i-1$,
\eqref{eq:G} becomes a geometric series,
\begin{align}
G(\Delta f)&=\bigg|\frac1N\sum_{i=1}^{N}e^{-j2\pi(i-1)\chi\Delta f/f_c}\bigg|^2=\bigg[\frac{\sin(\pi N\chi\Delta f/f_c)}{N\sin(\pi\chi\Delta f/f_c)}\bigg]^2,
\label{eq:dirproof}
\end{align}
which is \eqref{eq:dirichlet}. Over the main lobe
$N\sin(\pi\chi\Delta f/f_c)\approx N\pi\chi\Delta f/f_c$, so with $z=N\pi\chi\Delta f/f_c$ the
condition $G=1/2$ reads $\sin z/z=1/\sqrt2$, giving $z=z^\star$. Hence the half-power
half-width is $\Delta f_{3\mathrm{dB}}=z^\star f_c/(\pi N\chi)$ and
$B_{\mathrm{coh}}=2\Delta f_{3\mathrm{dB}}$ is \eqref{eq:bcoh}.
\end{IEEEproof}
Equation~\eqref{eq:dirichlet} is the array factor of a uniform linear array in the frequency variable $\Delta f/f_c$: the pinching array scans in frequency and its peak
walks off as $\Delta f$ grows, exactly beam squint \cite{dai2021delay,tarboush2021teramimo},
a phenomenon also motivating flexible-antenna remedies in phase-shifter arrays
\cite{sixdma2026}.

\begin{remark}\label{rem:novelty}
 In conventional wideband arrays, squint is a harmful artifact of
frequency-flat phase shifters, cured by TTD \cite{dai2021delay,tarboush2021teramimo}.
Squint in a PA is equally harmful, collapsing the gain away from $f_c$, but its cause
differs: alignment is set by physical path length, correct only modulo a wavelength
(Lemma~\ref{lem:cophase}), not by an electronic phase correct only at one frequency.
This is consequential: because co-phasing is an \emph{integer} condition on path
length, the admissible positions form an exact lattice (Theorem~\ref{thm:mip}), which
makes the placement problem exactly solvable and lets a few feed delays reach the
dispersion floor with $\mathcal{O}(\sqrt N)$ rather than $\mathcal{O}(N)$ elements.
Squint is thus the same effect in consequence, but a more structured object in
mechanism.
\end{remark}
\begin{remark}[Frequency-selective coherence loss vs.\ angular squint]\label{rem:terminology}
With a single fixed user and no beam scanning there is no angular walk-off: the effect here
is a frequency-selective loss of coherent combining gain, not a spatial beam that swings
with frequency. It is, however, the \emph{same} mechanism: if the array were steered, the coherent focal point would drift with frequency, so the Dirichlet-kernel description is exact rather than analogical, now in the frequency variable $\Delta f/f_c$ in place of a spatial angle. We retain the term ``beam squint'' for continuity with the wideband-array literature \cite{dai2021delay}, with this distinction understood.
\end{remark}

\begin{remark}[Coherent aperture vs.\ placement range]\label{rem:aperture}
The coherent cluster spans only $L=(N-1)\lambda_c/(\neff-\cos\zeta)$, which is
centimetre-scale at THz (e.g.\ $L\!\approx\!2$~cm for $N=29$ at $300$~GHz, $\neff=1.4$). The
metres-long waveguide is therefore used not to form a large coherent aperture but to
\emph{position} this small cluster at the pathloss-optimal site near the user, so the long
guide provides \emph{placement freedom}, not aperture. This smallness validates the constant-$\zeta$
linearization $u_i\approx-\cos\zeta\,v_i$ used throughout. A small $L$ simultaneously widens the
coherence bandwidth ($\Bcoh\propto1/L$) and the spatial robustness ($\Wsp\propto1/L$) at the
price of a proportionally smaller gain $N\propto L$, the $1/N$ law of
Proposition~\ref{prop:law}.
\end{remark}

Because the coherent array gain equals $N$ while the fractional coherence bandwidth is
$B_{\mathrm{coh}}/f_c=0.886/(N\chi)$, their product
\begin{equation}
N\cdot\frac{B_{\mathrm{coh}}}{f_c}=0.886\,\frac{n_{\mathrm{eff}}-\cos\zeta}{n_g-\cos\zeta}
\label{eq:gbw}
\end{equation}
is independent of $N$: array gain can be traded for coherence bandwidth only at a fixed
product set by the guide dispersion and the geometry. This invariance is one instance of a
single law that also governs the dispersion floor of Section~\ref{sec:gvd} and the spatial
robustness of Section~\ref{sec:opt}.

\begin{proposition}[Generalized gain--coherence law]\label{prop:law}
Let a size-$N$ consecutive-order cluster have, about a design point, a dominant residual
per-pinch phase of the single-term form
$\psi_i(\epsilon)=2\pi(i-1)\,c_p\,\epsilon^{\,p}$ in a scalar perturbation $\epsilon$, with
constant $c_p>0$ and integer order $p\ge1$. Then the normalized gain is the Dirichlet kernel
$G(\epsilon)=[\sin(\pi N c_p\epsilon^{p})/(N\sin(\pi c_p\epsilon^{p}))]^2$, its two-sided
$3$-dB half-width obeys $\epsilon_{3\mathrm{dB}}=(z^\star/\pi N c_p)^{1/p}\propto N^{-1/p}$,
and the gain--coherence product scales as $N\,\epsilon_{3\mathrm{dB}}\propto N^{\,1-1/p}$.
\end{proposition}
\begin{IEEEproof}
With $z=\pi N c_p\epsilon^{p}$ the summed geometric series gives the stated kernel; the
main-lobe half-power condition $\sin z/z=1/\sqrt2$ yields $z=z^\star$, i.e.\
$\epsilon^{p}=z^\star/(\pi N c_p)$, so $\epsilon_{3\mathrm{dB}}\propto N^{-1/p}$ and
$N\,\epsilon_{3\mathrm{dB}}\propto N^{1-1/p}$.
\end{IEEEproof}

\noindent Proposition~\ref{prop:law} unifies the three trade-offs through a common exponent governing their gain--bandwidth scaling. TTD correction thus does not merely displace the trade-off; it \emph{changes its order} ($p{=}1\!\to\!2$), converting the $N$-invariant ceiling into one that improves with array size, the sense in which the uncorrected array is $1/N$-limited, whereas the corrected array is only $1/\sqrt N$-limited.

\subsection{The Irreducible Dispersion Floor}\label{sec:gvd}
Theorem~\ref{thm:bcoh} retained only the first order in $\Delta f$, where the group index
merely rescaled the coherence bandwidth. That first-order squint is a pure delay effect, and
we show below that it is removable by TTD compensation. The genuinely dispersive
effect appears at second order, through group-velocity dispersion, and it cannot be removed
by any delay, so it sets a fundamental floor on the usable bandwidth. This section isolates
and quantifies that floor.

We expand the propagation constant about the carrier angular frequency $\omega_c=2\pi f_c$ to
second order. Since the free-space phase $\omega d_i/c$ is exactly linear in $\omega$ and
carries no curvature, only the guided term contributes, and \cite{balanis2012advanced,yeh2008essence}
\begin{equation}
\beta(\omega)=\beta_0+\beta_1\,\Delta\omega+\tfrac12\beta_2\,\Delta\omega^2
+\mathcal{O}(\Delta\omega^3),\qquad\Delta\omega=2\pi\Delta f,
\label{eq:betaexp}
\end{equation}
where $\beta_1=1/v_g=n_g/c$ is the group delay per unit length, $v_g$ is the group velocity,
and $\beta_2=\mathrm d^2\beta/\mathrm d\omega^2|_{\omega_c}$ is the group-velocity-dispersion
parameter, with units of $\mathrm{s^2/m}$. A non-dispersive guide has $\beta_2=0$.
\begin{lemma}\label{lem:secondorder}
Retaining \eqref{eq:betaexp}, the residual phase of pinch $i$ relative to pinch $1$
decomposes into a linear group-delay term and a quadratic dispersion term,
\begin{equation}
\psi_i(\Delta f)=2\pi\,\Delta f\,\tau_i+2\pi^2\beta_2\,v_i\,\Delta f^2,
\label{eq:psi2}
\end{equation}
with $\tau_i=(u_i+n_g v_i)/c$ as in Lemma~\ref{lem:groupdelay} and $v_i=x_i-x_1$.
\end{lemma}
\begin{IEEEproof}
The total phase is $\phi_i(\omega)=(\omega/c)d_i+\beta(\omega)x_i$. Only the guided term
carries curvature, so with \eqref{eq:betaexp} the difference relative to pinch $1$, after
subtracting the co-phasing constant $2\pi m_i$, is
$\psi_i(\Delta f)=\Delta\omega[u_i/c+\beta_1 v_i]+\tfrac12\beta_2\Delta\omega^2 v_i$. The
first bracket equals $\tau_i$ since $\beta_1=n_g/c$, and with $\Delta\omega=2\pi\Delta f$ the
two terms become \eqref{eq:psi2}.
\end{IEEEproof}

The two terms are physically distinct. A frequency-flat feed delay $\tau_i^{\mathrm{TTD}}$
adds a phase $-2\pi f\tau_i^{\mathrm{TTD}}$ that is linear in $\Delta f$, so it cancels the
group-delay term exactly when $\tau_i^{\mathrm{TTD}}=\tau_i$. No delay can reproduce a term
in $\Delta f^2$, so the dispersion term is irreducible. After ideal first-order compensation
the residual phase is purely
\begin{equation}
\psi_i^{\mathrm{res}}(\Delta f)=2\pi^2\beta_2\,v_i\,\Delta f^2.
\label{eq:psires}
\end{equation}
\begin{theorem}\label{thm:floor}
Under the canonical placement, for which $v_i=(i-1)\lambda_c/(n_{\mathrm{eff}}-\cos\zeta)$, and
with ideal cancellation of the first-order squint, the residual array gain is a Dirichlet
kernel quadratic in frequency,
\begin{equation}
G_{\mathrm{res}}(\Delta f)=\bigg[\frac{\sin(N\mu/2)}{N\sin(\mu/2)}\bigg]^2,\qquad\!\!\!\!
\mu=2\pi^2\beta_2\,\frac{\lambda_c}{n_{\mathrm{eff}}-\cos\zeta}\,\Delta f^2,
\label{eq:Gres}
\end{equation}
and the two-sided $3$-dB dispersion-limited bandwidth is
\begin{equation}
B_{\mathrm{res}}=\frac{2}{\pi}\sqrt{\frac{z^\star(n_{\mathrm{eff}}-\cos\zeta)f_c}{N\beta_2 c}}
\approx0.751\sqrt{\frac{(n_{\mathrm{eff}}-\cos\zeta)f_c}{N\beta_2 c}}.
\label{eq:Bres}
\end{equation}
\end{theorem}
\begin{IEEEproof}
Ideal compensation leaves \eqref{eq:psires}. Substituting
$v_i=(i-1)\lambda_c/(n_{\mathrm{eff}}-\cos\zeta)$ gives $\psi_i^{\mathrm{res}}=(i-1)\mu$, and
summing the geometric series as in the proof of Theorem~\ref{thm:bcoh} yields \eqref{eq:Gres}.
The main-lobe condition $N\mu/2=z^\star$ solves to
$\Delta f_{\mathrm{res}}=\sqrt{z^\star(n_{\mathrm{eff}}-\cos\zeta)/(N\pi^2\beta_2\lambda_c)}$,
and $B_{\mathrm{res}}=2\Delta f_{\mathrm{res}}$ with $\lambda_c=c/f_c$ is \eqref{eq:Bres}.
\end{IEEEproof}

\begin{remark}[Interpreting the dispersion floor]\label{rem:floor}
Two features of \eqref{eq:Bres} are important. First, the floor scales as
$B_{\mathrm{res}}\propto1/\sqrt N$, whereas the uncorrected coherence bandwidth of
\eqref{eq:bcoh} scales as $1/N$. TTD correction therefore improves the
gain-bandwidth scaling from $1/N$ to $1/\sqrt N$, and the floor widens as $1/\sqrt{\beta_2}$
for a less dispersive guide. Second, the dispersion term is masked by the first-order squint
until the latter is removed. Equating the two terms of \eqref{eq:psi2} at broadside gives the
crossover offset $\Delta f^\star=n_g/(\pi\beta_2 c)$, below which the linear squint dominates
and dispersion is negligible. Since the uncorrected coherence bandwidth is in practice far
smaller than $\Delta f^\star$, waveguide dispersion is immaterial to the uncorrected array and
becomes the binding limit only once the first-order squint has been compensated, which is the
subject of Section~\ref{sec:fix}. Finally, $\beta_2$ combines material dispersion, from the
frequency dependence of the dielectric index, and modal dispersion, from the guided-mode
geometry, the latter dominating near modal cutoff \cite{yeh2008essence,balanis2012advanced}.
Both enter \eqref{eq:psires} and \eqref{eq:Bres} identically, and $\beta_2$ is precisely the
quantity discarded by the lossless narrowband models, so by \eqref{eq:Bres} it alone sets the
ultimate wideband limit of a pinching array.
\end{remark}
\section{Wideband Achievable Rate under Random User Location}\label{sec:rate}
We now relate the squint and dispersion results to the ergodic spectral efficiency of
\eqref{eq:rate}, averaged over the band and over a uniformly random user location with
$x_m\sim\mathcal U[0,D_x]$ and $y_m\sim\mathcal U[-D_y/2,D_y/2]$. The main finding is
structural. The user-position statistics and the wideband squint separate, and the squint
enters the rate only through a single deterministic scalar.

A first observation reduces the wideband rate to known narrowband quantities. By the
linearity of expectation applied to \eqref{eq:rate},
\begin{equation}
\bar R=\frac1K\sum_{k=0}^{K-1}R_k,\qquad R_k=\mathbb{E}_{x_m,y_m}\big[\log_2(1+\gamma_k)\big],
\label{eq:persub}
\end{equation}
where each $R_k$ is the narrowband single-pinch ergodic rate evaluated at $f_k$ with the
received SNR scaled by the squint factor $G(\Delta f_k)$. The wideband rate is thus the
subcarrier average of the closed-form narrowband expression derived for loss-aware pinching
links in \cite{tyrovolas2026performance}, and it inherits that machinery term by term.
Equation~\eqref{eq:persub} is exact at every SNR. The low-SNR regime, which is the relevant
one for power- or distance-limited THz links \cite{rappaport2019wireless},
admits a fully decoupled closed form.
\begin{theorem}\label{thm:lowsnr}
In the low-SNR regime $\gamma_k\ll1$, the wideband ergodic rate factors as
\begin{equation}
\bar R\approx\frac{\eta_{\mathrm{wb}}}{\ln2}\,\mathbb{E}[\gamma_0],
\label{eq:Rlow}
\end{equation}
where the wideband efficiency
\begin{equation}
\eta_{\mathrm{wb}}=\frac1K\sum_{k=0}^{K-1}\Big(\frac{f_c}{f_k}\Big)^2 G(\Delta f_k)\in(0,1]
\label{eq:etawb}
\end{equation}
is deterministic and captures jointly the $1/f^2$ roll-off and the squint, and the
position-averaged reference SNR is available in closed form as
\begin{equation}
\mathbb{E}[\gamma_0]=\frac{PN}{K\sigma^2}\,\eta(f_c)\,
\frac{1-e^{-\alpha D_x}}{\alpha D_x}\,\frac{2}{hD_y}\arctan\!\frac{D_y}{2h}.
\label{eq:Egamma0}
\end{equation}
\end{theorem}
\begin{IEEEproof}
For $\gamma_k\ll1$, $\log_2(1+\gamma_k)\approx\gamma_k/\ln2$, so
$\bar R\approx(1/K\ln2)\sum_k\mathbb{E}[\gamma_k]$. Write $\gamma_k=\gamma_0\rho_k$ with
reference SNR $\gamma_0=(PN/K\sigma^2)\eta(f_c)e^{-\alpha x_m}/(y_m^2+h^2)$ and
$\rho_k=[\eta(f_k)/\eta(f_c)]e^{-[\alpha(f_k)-\alpha(f_c)]x_m}G(\Delta f_k)$. The attenuation
dispersion over the band is negligible, so $\rho_k\approx(f_c/f_k)^2 G(\Delta f_k)$ is
independent of position, and $\mathbb{E}[\gamma_k]=\rho_k\mathbb{E}[\gamma_0]$. Summing gives
\eqref{eq:Rlow} with $\eta_{\mathrm{wb}}=(1/K)\sum_k\rho_k$. Since $x_m$ and $y_m$ are
independent, $\mathbb{E}[\gamma_0]=(PN/K\sigma^2)\eta(f_c)\,\mathbb{E}[e^{-\alpha x_m}]\,
\mathbb{E}[1/(y_m^2+h^2)]$. The elementary integrals
$\mathbb{E}[e^{-\alpha x_m}]=(1-e^{-\alpha D_x})/(\alpha D_x)$ and
$\mathbb{E}[1/(y_m^2+h^2)]=(2/hD_y)\arctan(D_y/2h)$ give \eqref{eq:Egamma0}.
\end{IEEEproof}

\begin{remark}[Validity of the attenuation decoupling]\label{rem:alphacaveat}
The step $\rho_k\approx(f_c/f_k)^2 G(\Delta f_k)$ drops the attenuation dispersion by
assuming $\alpha(f_k)\approx\alpha(f_c)$. From \eqref{eq:alpha}, the excess exponent across
the band is $[\alpha(f_k)-\alpha_0]x_m\approx\alpha_0\,s\,(\Delta f_k/f_c)x_m$, so the
decoupling requires $s\,(B/2f_c)\,\alpha_0 D_x\ll1$. For large loss exponent $s$, large
fractional bandwidth, or long guide runs this fails; the factor then reacquires a
deterministic, position-coupled dependence, the amplitude-dispersion analogue of the
molecular-absorption taper of Section~\ref{subsec:absorption}, and the factorization
\eqref{eq:Rlow} holds only with a position-averaged correction.
\end{remark}

Equation~\eqref{eq:Rlow} isolates two independent effects. The user-position statistics are
carried by $\mathbb{E}[\gamma_0]$, whose closed form in \eqref{eq:Egamma0} reduces to the
narrowband result of \cite{tyrovolas2026performance} when evaluated at a single frequency.
The wideband squint is carried by the scalar $\eta_{\mathrm{wb}}$. As the bandwidth shrinks
relative to the coherence bandwidth of Theorem~\ref{thm:bcoh}, every $G(\Delta f_k)\to1$ and
$\eta_{\mathrm{wb}}\to1$, so the narrowband rate is recovered. As the bandwidth grows past
$B_{\mathrm{coh}}$, an increasing fraction of subcarriers fall below the $3$-dB point and
$\eta_{\mathrm{wb}}$ decreases, so the fractional loss $1-\eta_{\mathrm{wb}}$ is a direct
closed-form measure of the squint penalty. The deterministic effect of molecular absorption,
significant over THz distances \cite{tarboush2021teramimo}, is
developed next; at general SNR, \eqref{eq:persub} provides the exact rate as the subcarrier
average of the narrowband closed form.

\subsection{Molecular Absorption and the Squint}\label{subsec:absorption}
With molecular absorption the free-space factor becomes $\eta(f)\!\to\!\eta(f)e^{-\kappa(f)d}$,
where $\kappa(f)$ [Np/m] is the power absorption coefficient and $d$ the free-space hop. The
pinch amplitude in \eqref{eq:Hwi} acquires the extra factor $e^{-\kappa(f)d_i/2}$. For the
small cluster of Remark~\ref{rem:aperture} ($d_i\!\approx\!d$, so
$\kappa(f)(d_i-d)\!\ll\!1$), this factor is \emph{common-mode}: it leaves the array factor
$G$ in \eqref{eq:G} unchanged and instead reweights each subcarrier. Hence the wideband
efficiency \eqref{eq:etawb} generalizes to
\begin{equation}
\etawb^{\mathrm{abs}}=\frac1K\sum_{k=0}^{K-1}
\Big(\frac{f_c}{f_k}\Big)^{2}e^{-[\kappa(f_k)-\kappa(f_c)]\,d}\;G(\Delta f_k).
\label{eq:etawbabs}
\end{equation}
Absorption and squint therefore \emph{multiply}: absorption is a deterministic amplitude
taper over the subcarriers, the squint a phase-driven coherence loss. Absorption can depress
$\etawb$ but does not interact with the squint beyond reweighting.

Inside a transmission window (between water-vapour lines) $\kappa$ is smooth; expanding about
$f_c$, $\kappa(f)\approx\kappa_0+\kappa_1\Delta f+\tfrac12\kappa_2\Delta f^2$, the slope
$\kappa_1$ produces an asymmetric spectral tilt $e^{-\kappa_1 d\,\Delta f}$ (shifting the
effective passband centroid and lowering $\etawb$) and the curvature $\kappa_2$ a symmetric
roll-off $e^{-\frac12\kappa_2 d\,\Delta f^2}$. Defining the two-sided $3$-dB
\emph{absorption-limited bandwidth} from the symmetric term,
\begin{equation}
\Babs=2\sqrt{\frac{2\ln2}{\kappa_2\,d}}=\frac{2.355}{\sqrt{\kappa_2\,d}},
\label{eq:Babs}
\end{equation}
the usable band is the intersection of the squint- and absorption-limits, mirroring~\eqref{eq:buse}:
\begin{equation}
\Buse=\min\{\,\Bcoh\ \text{(or }\Bres\text{ after TTD)},\ \Babs\,\}.
\label{eq:buseabs}
\end{equation}
Which limit binds is distance-dependent: equating \eqref{eq:Babs} with
$\Bcoh=0.886f_c/(N\chi)$ gives the crossover hop
\begin{equation}
d^\star=\frac{1}{\kappa_2}\Big(\frac{2.659\,N\chi}{f_c}\Big)^{2},
\label{eq:dstar}
\end{equation}
so squint binds for $d<d^\star$ and absorption for $d>d^\star$. Over the short LOS hops that PASS creates, absorption typically does not dominate; over longer hops or near a
window edge it does, and \eqref{eq:dstar} marks the boundary. Near a water absorption line at $f_\ell$, $\kappa(f)$ varies sharply across the band, causing $e^{-\kappa(f)d}$ to suppress the affected subcarriers. Consequently, $\etawb^{\mathrm{abs}}$ must be evaluated numerically using spectroscopic line databases. As in practical THz systems, the design response is to confine the transmission band to a low-absorption window, where \eqref{eq:Babs}--\eqref{eq:dstar} apply.

\subsection{Rate-Optimal Pinch Count}\label{subsec:Nstar}
Theorem~\ref{thm:bcoh} shows coherence shrinks as $\Bcoh\propto1/N$ while the coherent SNR
grows as $\gamma_0\propto N$ by \eqref{eq:snr}, so band-integrated rate has an interior sweet
spot. Let $\gamma_0=Ng$ denote the coherent band-centre SNR, where $g=\gamma_0/N$ is the
single-pinch (full-power) reference SNR; the out-of-band subcarriers combine incoherently
($G\!\approx\!1/N$ on average) and fall back to this same $g$, independent of $N$. Adopt the
sharp-coherence idealization $G(\Delta f)\approx\mathbf 1\{|\Delta f|\le \Bcoh(N)/2\}$.
\begin{proposition}[Rate-optimal pinch count]\label{prop:Nstar}
Under the sharp-coherence idealization and neglecting the slow roll-off, the band-averaged
rate $\bar R(N)=\tfrac1B\int_{-B/2}^{B/2}\log_2(1+Ng\,G(\Delta f))\,d(\Delta f)$ is strictly
increasing for $N<N^\star$ and strictly decreasing for $N>N^\star$, with the unique maximizer
\begin{equation}
N^\star=\frac{0.886\,f_c}{\chi\,B}\qquad\Longleftrightarrow\qquad\Bcoh(N^\star)=B.
\label{eq:Nstar}
\end{equation}
The optimum places the coherence bandwidth exactly at the signal bandwidth.
\end{proposition}
\begin{IEEEproof}
If $N\le N^\star$ then $\Bcoh(N)\ge B$, the whole band is coherent, every subcarrier carries
SNR $Ng$, and $\bar R(N)=\log_2(1+Ng)$, strictly increasing. If $N>N^\star$ then a fraction
$\Bcoh(N)/B=N^\star/N$ of the band is coherent and the rest contributes zero under the
idealization, so $\bar R(N)=N^\star f(N)$ with $f(N)\triangleq\log_2(1+Ng)/N$. Then
$f'(N)=[N^2\ln2]^{-1}[Ng/(1+Ng)-\ln(1+Ng)]=-\phi(Ng)/(N^2\ln2)$, where
$\phi(u)=\ln(1+u)-u/(1+u)$ satisfies $\phi(0)=0$ and $\phi'(u)=u/(1+u)^2>0$, hence
$\phi(u)>0$ and $f'(N)<0$ for $N>N^\star$. Continuity at $N^\star$ completes the proof.
\end{IEEEproof}
\begin{corollary}[TTD-corrected sweet spot]\label{cor:Nttd}
Replacing $\Bcoh(N)$ by the dispersion floor $\Bres(N)$ of Theorem~\ref{thm:floor}, the
diminishing-returns knee moves to
$N^\star_{\mathrm{TTD}}=0.564\,(\neff-\cos\zeta)f_c/(\beta_2 c B^2)$ (from
$\Bres(N^\star_{\mathrm{TTD}})=B$), with
$N^\star_{\mathrm{TTD}}/N^\star=0.637(\ngr-\cos\zeta)/(\beta_2 c B)\gg1$: TTD
correction lifts the useful pinch count, hence attainable array gain, by orders of magnitude.
\end{corollary}
\begin{remark}[Low-SNR knee vs.\ high-SNR peak]\label{rem:knee}
At strict low SNR, the decrease beyond $N^\star$ is $\mathcal O(g^2)$, so $\bar R$ remains essentially flat, yielding a diminishing-returns knee rather than a sharp peak. Restoring the incoherent floor shifts the maximizer modestly above $N^\star$. Nevertheless, $N^\star$ marks the onset of diminishing returns and remains the appropriate design target.
\end{remark}
\noindent At broadside $\chi=\ngr/\neff$, so \eqref{eq:Nstar} specializes to
$N^\star=0.886\,(\neff/\ngr)\,f_c/B$, scaling inversely with the signal
bandwidth. With $\beta_2\!\sim\!1$~ps$^2$/m the dispersion-limited count
$N^\star_{\mathrm{TTD}}$ of Corollary~\ref{cor:Nttd} is orders of magnitude larger, so post-TTD the binding constraint is spacing/aperture feasibility, not dispersion.

\section{Delay-Phase Squint Mitigation}\label{sec:fix}
Theorem~\ref{thm:floor} showed that ideal TTD correction leaves only the
dispersion floor $B_{\mathrm{res}}$. We now give a realizable architecture that approaches
this floor with few delay elements, by letting pinch positioning rather than radio-frequency
hardware perform the pathloss minimization and the fine alignment, while a small number of
feed delays remove the coarse squint. The design is the pinching counterpart of
sub-connected delay-phase precoding for wideband THz arrays \cite{dai2021delay}.

The array uses the multi-waveguide model of Section~\ref{subsec:geom}. Waveguide $w$ carries
$N_w$ pinches clustered about a site $\bar x_w$ and is driven through a feed delay $\tau_w$,
so the combined channel is \eqref{eq:Hcomb}. The per-pinch group delay and the
cluster-centroid delay are
\begin{equation}
\tau_{w,i}=\frac{d_{w,i}+n_g x_{w,i}}{c},\qquad\bar\tau_w=\frac1{N_w}\sum_{i=1}^{N_w}\tau_{w,i}.
\label{eq:tauwi}
\end{equation}
\begin{proposition}\label{prop:ttd}
The causal feed delays $\tau_w=\tau_{\max}-\bar\tau_w$ with $\tau_{\max}=\max_w\bar\tau_w$
cancel the inter-cluster group-delay differences to first order in $\Delta f$, leaving only
the intra-cluster group-delay spread and the common dispersion curvature.
\end{proposition}
\begin{IEEEproof}
The feed delay adds the frequency-flat phase $-2\pi f\tau_w$, a pure delay $\tau_w$ common to
the cluster. The effective group delay of pinch $(w,i)$ is
$\tau_{w,i}+\tau_w=(\tau_{w,i}-\bar\tau_w)+\tau_{\max}$. The global constant $\tau_{\max}$
does not affect relative phases, and the centroid delays $\bar\tau_w$, which carry the coarse
differences, are removed. Only the intra-cluster deviations $\tau_{w,i}-\bar\tau_w$ remain,
and by Lemma~\ref{lem:secondorder} the dispersion phase is unchanged by a frequency-flat
delay.
\end{IEEEproof}

With equal clusters $N_w=N/W$, each co-phased in the canonical sense with $N_w\gtrsim4$, two
residuals remain after Proposition~\ref{prop:ttd}. The intra-cluster first-order squint of a
size-$N_w$ cluster contributes a $3$-dB bandwidth $0.886(f_c/N_w)(n_{\mathrm{eff}}/n_g)$ by
Theorem~\ref{thm:bcoh}, and the dispersion term, unaffected by the feed delays and acting
across the full $N$-pinch aperture, contributes the floor $B_{\mathrm{res}}$ of
Theorem~\ref{thm:floor}. The gain falls below $1/2$ once either residual reaches its $3$-dB
point, so the usable bandwidth is
\begin{equation}
B_{\mathrm{use}}=\min\!\Big\{0.886\,\frac{f_c W}{N}\,\frac{n_{\mathrm{eff}}}{n_g},\
B_{\mathrm{res}}\Big\}.
\label{eq:buse}
\end{equation}
More feed delays widen the first term, which scales as $W/N$, until it meets the dispersion
floor, beyond which further delays bring no gain. Equating the two terms of \eqref{eq:buse}
and substituting $B_{\mathrm{res}}$ from \eqref{eq:Bres} at broadside gives
$W=B_{\mathrm{res}}Nn_g/(0.886 f_c n_{\mathrm{eff}})$, which simplifies to
\begin{equation}
W_{\min}\approx0.848\,\frac{n_g}{n_{\mathrm{eff}}}\sqrt{\frac{N\,n_{\mathrm{eff}}}
{\beta_2\,c\,f_c}}=\mathcal O(\sqrt N),
\label{eq:wmin}
\end{equation}
valid when the result satisfies $N/W_{\min}\gtrsim4$. The dispersion floor is therefore
attained with $\mathcal O(\sqrt N)$ feed delays rather than one per pinch, because pinch
positioning performs the fine alignment and the pathloss minimization at no delay cost. This
contrasts with the $\mathcal O(N)$ delay elements of a conventional delay-phase array. Setting $W=N$, one pinch per waveguide, makes the first term vanish and
recovers the pure floor $B_{\mathrm{res}}$, so \eqref{eq:wmin} shows that choice to be
wasteful. The feed delays follow in closed form from Proposition~\ref{prop:ttd}, while the
pinch positions are obtained from the placement problem of Section~\ref{sec:opt}.

\section{Optimal and Robust Pinch Placement}\label{sec:opt}
Sections~\ref{sec:both}--\ref{sec:fix} assumed the canonical placement. We now solve the
placement problem \eqref{eq:Theta}, prove the canonical choice optimal, analyze robustness to
an uncertain user position, and assemble these facts into the proposed algorithm with its
benchmarks.

\subsection{Problem and Exact Integer Reformulation}\label{subsec:mip}
The placement problem maximizes the band-averaged received power \eqref{eq:Theta} over pinch
positions and feed delays,
\begin{equation}
\begin{aligned}
(\mathrm P)\quad \max_{\{x_i\},\{\tau_w\}}\ &\ \Theta\\
\text{s.t.}\quad  x_i\in[0,D_x],\ &|x_i-x_j|\ge\Delta_{\min},\ i\neq j,\ \tau_w\ge0.
\end{aligned}
\label{eq:P}
\end{equation}
By Theorem~\ref{thm:lowsnr}, placement enters the ergodic rate only through the
band-averaged gain, and $\Theta$ \eqref{eq:Theta} is exactly that quantity
($\Theta\propto\sum_k G(\Delta f_k)$ under Assumption~\ref{as:amp}). Maximizing
$\Theta$ is thus no surrogate: the rate analysis sets this objective and fixes the
array size $N^\star$ (Proposition~\ref{prop:Nstar}) that placement realizes.
This problem is severely non-convex. Along any single coordinate $x_i$, the cross terms of
$|H|^2$ oscillate with the guided phase $(2\pi f_k/c)n_{\mathrm{eff}}x_i$, which advances by
$n_{\mathrm{eff}}D_x/\lambda_c$ cycles as $x_i$ sweeps the region. Each cycle creates a
maximum and a minimum, so the objective has on the order of $n_{\mathrm{eff}}D_x/\lambda_c$
stationary points per coordinate and up to $\mathcal O((n_{\mathrm{eff}}D_x/\lambda_c)^N)$
joint local optima. At $f_c=300$~GHz with $D_x=10$~m this is about $1.4\times10^4$ stationary
points per coordinate, the pinching analogue of the oscillatory landscape that forces grid
search in wideband array design \cite{bereyhi2025mimopass}. Gradient and Newton methods are
therefore unreliable, and we instead exploit an exact integer structure.

The structure follows from the monotonicity of the electrical path. Since
$P(x)=d(x)+n_{\mathrm{eff}}x$ has derivative
$P'(x)=n_{\mathrm{eff}}-\cos\zeta(x)\ge n_{\mathrm{eff}}-1>0$, it is a strictly increasing
bijection on $[0,D_x]$. Fixing a reference path $P_0$, the co-phased lattice points
$x^{(0)}(m)$ defined by $P(x^{(0)}(m))=P_0+m\lambda_c$ for $m\in\mathbb{Z}$ are the positions
at which a pinch is co-phased with the reference at $f_c$. Every position decomposes uniquely
as
\begin{equation}
x_i=x^{(0)}(m_i)+\delta_i,\qquad m_i\in\mathbb{Z},\ \ \delta_i\in(-\ell/2,\ell/2],
\label{eq:coarsefine}
\end{equation}
into an integer order $m_i$ and a sub-wavelength offset $\delta_i$, where
$\ell=\lambda_c/(n_{\mathrm{eff}}-\cos\zeta)$ is the local cell width.
\begin{theorem}\label{thm:mip}
The placement problem \eqref{eq:P} is equivalent, through the change of variables
\eqref{eq:coarsefine}, to the mixed-integer program
\begin{align}
(\mathrm P')\quad&\max_{\{m_i\}\in\mathbb{Z}^N,\{\delta_i\},\{\tau_w\}}\ \Theta
\notag \\\quad\text{s.t.} &\quad\eqref{eq:coarsefine}\ \text{and the constraints of }(\mathrm P).
\label{eq:Pprime}
\end{align}
The equivalence is exact. The combinatorial difficulty resides entirely in the integer orders
$\{m_i\}$, which set the wideband array factor of Theorem~\ref{thm:bcoh}, while for fixed
orders the objective is smooth in $\{\delta_i\}$ within each cell, with order-one stationary
points per coordinate.
\end{theorem}
\begin{IEEEproof}
Monotonicity of $P$ makes the map $x\mapsto(m,\delta)$ a bijection between $[0,D_x]$ and the
lattice cells, so the change of variables \eqref{eq:coarsefine} preserves the feasible set and
the objective without approximation. Fixing $\{m_i\}$ fixes the integer part of every
electrical path, hence the co-phasing pattern that generates the array factor, while the
dependence on $\delta_i$ enters only through the amplitude and a sub-wavelength phase that
sweeps less than one cycle over a cell, which is smooth.
\end{IEEEproof}

This reformulation is specific to pinching. It follows from Lemma~\ref{lem:cophase}, where
co-phasing is an equality on path length modulo $\lambda_c$, which makes the admissible
positions an exact integer lattice. A conventional phased array, whose element phases are set
by electronic shifters rather than by physical path, has no such lattice. Moreover, with the
optimal feed delays of Proposition~\ref{prop:ttd} the objective separates across clusters,
since distinct clusters are separated by many wavelengths and their cross terms average to
zero over the band, so the order and offset selection is solved independently within each
cluster.

\subsection{Tractable Solution}\label{subsec:solve}
The integer layer, which appears combinatorial, has a closed-form solution in the operating
regime.
\begin{proposition}\label{prop:consec}
For $B\le f_c/[(N_w-1)q^\star]$, that is throughout the coherent regime, the band-averaged
gain of a size-$N_w$ cluster is maximized over all feasible integer order sets (adjacent
order gaps at least $q^\star$ of Section~\ref{subsec:feas}) by the minimal-step progression
$m_i=m_0+(i-1)q^\star$, uniquely up to a global shift. When the spacing-feasibility condition
\eqref{eq:feas} holds, $q^\star=1$ and the canonical placement of
Sections~\ref{sec:both}--\ref{sec:fix} is optimal.
\end{proposition}
\begin{IEEEproof}
Neglecting the slow roll-off, the band-averaged gain is proportional to
$\int_{-\Omega/2}^{\Omega/2}|\mathrm{AF}(\omega)|^2\,d\omega$ with
$\mathrm{AF}(\omega)=(1/N_w)\sum_i e^{-j\omega m_i}$ and $\Omega=2\pi B/f_c$. Expanding the
square,
\begin{equation}
\int_{-\Omega/2}^{\Omega/2}\!\!|\mathrm{AF}|^2\,d\omega
=\frac{\Omega}{N_w^2}\sum_{i,j}\sinc\!\Big(\frac{(m_i-m_j)\Omega}{2}\Big)
\triangleq\frac{\Omega}{N_w^2}\,S(\mathbf m),
\label{eq:consecproof}
\end{equation}
with $\sinc(x)=\sin x/x$; maximizing the band-averaged gain is equivalent to maximizing
$S(\mathbf m)$ over feasible integer sets. Order $m_1<\dots<m_{N_w}$ with gaps
$g_r=m_{r+1}-m_r\ge q^\star$. The regime condition keeps every pairwise argument of the
minimal-step configuration in $(0,\pi)$, where $\sinc$ is positive and strictly decreasing.
Suppose some gap $g_r\ge q^\star+1$; form $\mathbf m'$ by decrementing every $m_i$, $i>r$, by
one. Pairs on the same side of the cut keep their differences; each crossing pair $i\le r<j$
has its difference reduced by exactly one, remaining $\ge q^\star$. Since the reduction only decreases pairwise differences toward those of the minimal-step configuration, all arguments stay in $(0,\pi)$, so every crossing term strictly increases and none decreases: $S(\mathbf m')>S(\mathbf m)$. Iterating removes all excess gaps, so the minimal-step progression, the unique gap-floored configuration up to a global shift, is the unique maximizer.
\end{IEEEproof}

With the orders fixed to the minimal-step progression, only the sub-wavelength offsets
$\{\delta_i\}$ remain, and the objective is smooth in them by Theorem~\ref{thm:mip}. Under the
equal-amplitude approximation of Assumption~\ref{as:amp} these offsets vanish at the optimum.
They do work only when the amplitude varies across the cluster, as for a large aperture or a
strong attenuation gradient, where, with the orders fixed, a successive convex approximation
adapted from the movable-antenna capacity framework \cite{ma2024mimo} refines them through the
update
\begin{equation}
\{\delta_i^{(t+1)}\}=\arg\max_{\{\delta_i\}}\ \widetilde\Theta\quad\text{s.t.}\quad
\text{spacing and }|\delta_i|\le\ell/2,
\label{eq:scaupdate}
\end{equation}
a quadratic program with a projected closed-form solution \cite{boyd2004convex} whose
surrogate lower-bounds the objective and is tight at the iterate, so the iterates increase
monotonically to a stationary point \cite{razaviyayn2013unified}. The per-cluster cost is
$\mathcal O(I_{\mathrm{SCA}}(N_wK+N_w^{1.5}))$ with $I_{\mathrm{SCA}}$ the iteration count and
$N_w^{1.5}$ the interior-point quadratic program, and across clusters the subproblems are
independent with closed-form feed delays, giving an overall
$\mathcal O(W I_{\mathrm{SCA}}(N_wK+N_w^{1.5}))$, polynomial in $N$, $W$, and $K$. The
exponential landscape of $(\mathrm P)$ is thus eliminated, the integer layer being closed-form
and the continuous layer smooth.

\subsection{Spacing Feasibility and Order Skipping}\label{subsec:feas}
The minimal-step progression of Proposition~\ref{prop:consec} assumes consecutive orders are
feasible. With consecutive orders, adjacent pinches sit one cell apart,
\begin{equation}
\Delta x=x_{i+1}-x_i=\ell=\frac{\lambda_c}{\neff-\cos\zeta}.
\label{eq:cell}
\end{equation}
At $f_c=300$~GHz, broadside, $\neff=1.4$: $\ell\approx0.71\lambda_c\approx0.71$~mm, which is
\emph{sub-wavelength}. If the coupling-limited spacing $\Delta_{\min}$ of \eqref{eq:P} exceeds
$\ell$, consecutive placement is infeasible. The feasibility condition is
\begin{equation}
\Delta_{\min}\le\ell=\frac{\lambda_c}{\neff-\cos\zeta}.
\label{eq:feas}
\end{equation}
When \eqref{eq:feas} fails, co-phasing must skip orders: use step
\begin{equation}
q^\star=\Big\lceil\frac{\Delta_{\min}}{\ell}\Big\rceil
=\Big\lceil\frac{\Delta_{\min}(\neff-\cos\zeta)}{\lambda_c}\Big\rceil\ge1,
\label{eq:qstar}
\end{equation}
giving adjacent spacing $q^\star\ell\ge\Delta_{\min}$. The residual phase of
Theorem~\ref{thm:bcoh} becomes $\psi_i=2\pi(i-1)q^\star\chi\,\Delta f/f_c$, i.e.\
$\chi\!\to\!q^\star\chi$, so the coherence bandwidth and the spatial robustness of
Theorem~\ref{thm:spatial} both shrink by $q^\star$:
\begin{equation}
\Bcoh^{(q)}=\frac{\Bcoh}{q^\star},\qquad \Wsp^{(q)}=\frac{\Wsp}{q^\star}.
\label{eq:qscaling}
\end{equation}
Both are aperture effects: order-skipping stretches the aperture to
$L^{(q)}\!\approx\!q^\star L$, and a larger aperture means more delay spread (narrower
$\Bcoh$) and tighter focus (narrower $\Wsp$), consistent with Proposition~\ref{prop:law}.
When \eqref{eq:feas} holds, $q^\star=1$ and Theorems~\ref{thm:bcoh} and~\ref{thm:spatial} hold
verbatim.

\subsection{Robust Placement under Position Uncertainty}\label{subsec:robust}
The previous subsections optimized placement for a known user position. In practice the
position is known only to lie in an uncertainty set $\mathcal U$, such as a localization cell.
Whereas Section~\ref{sec:rate} averaged over the user distribution, we seek here a worst-case
guarantee,
\begin{equation}
(\mathrm P_{\mathrm{rob}})\quad\max_{\{x_i\},\{\tau_w\}}\
\min_{(x_m,y_m)\in\mathcal U}\ \Theta.
\label{eq:Prob}
\end{equation}
Robustness is governed by a spatial coherence region that mirrors the coherence bandwidth of
Theorem~\ref{thm:bcoh}.
\begin{theorem}\label{thm:spatial}
For a size-$N$ consecutive-order cluster co-phased for a reference user at distance $d$, the
array gain toward a user displaced axially by $s$ is the spatial Dirichlet kernel
\begin{equation}
G_s(s)=\bigg[\frac{\sin(\pi N s/(n_{\mathrm{eff}}d))}{N\sin(\pi s/(n_{\mathrm{eff}}d))}\bigg]^2,
\label{eq:Gspatial}
\end{equation}
whose two-sided $3$-dB position-coherence width is
\begin{equation}
W_{\mathrm{spatial}}=0.886\,\frac{n_{\mathrm{eff}}\,d}{N}=0.886\,\frac{\lambda_c\,d}{L},
\label{eq:Wspatial}
\end{equation}
with $L=(N-1)\lambda_c/n_{\mathrm{eff}}$ the cluster aperture.
\end{theorem}
\begin{IEEEproof}
Let the pinches sit at $x_i=\bar x+\xi_i$ with $\xi_i=(i-(N-1)/2)\lambda_c/n_{\mathrm{eff}}$,
co-phased for the reference user at distance $d$. For an axial displacement $s$, the distance
perturbs as $d_i(s)-d_i(0)\approx-s\xi_i/d$ to first order. The co-phasing cancels the
reference phase, so the array sum is $(1/N)\sum_i e^{jk_c s\xi_i/d}$ with $k_c=2\pi/\lambda_c$.
The per-index phase is $2\pi s/(n_{\mathrm{eff}}d)$, a geometric series whose squared magnitude
is \eqref{eq:Gspatial}. The main-lobe condition $\pi N s/(n_{\mathrm{eff}}d)=z^\star$ gives the
half-width, and doubling it yields \eqref{eq:Wspatial}.
\end{IEEEproof}

Because clusters are millimeter-scale, $W_{\mathrm{spatial}}$ is of order decimeters to
meters, so a pinching cluster is intrinsically robust to position uncertainty within its focal
spot. The gain-robustness product $N\cdot W_{\mathrm{spatial}}/d=0.886\,n_{\mathrm{eff}}$ is
independent of $N$, so array gain trades against position robustness exactly as it trades
against bandwidth in \eqref{eq:gbw}, the first-order ($p=1$) case of
Proposition~\ref{prop:law}. The robust problem then has a simple structure.
\begin{proposition}\label{prop:robust}
Let $\mathcal U$ be a convex cell with axial extent $[x_{\mathrm{lo}},x_{\mathrm{hi}}]$. The
worst-case user in \eqref{eq:Prob} lies on the boundary of $\mathcal U$. If
$x_{\mathrm{hi}}-x_{\mathrm{lo}}\le W_{\mathrm{spatial}}$, the focal-spot-centered placement is
robust, with worst-case gain within $3$~dB of nominal. Otherwise the robust cluster center
solves the one-dimensional problem
\begin{equation}
\bar x^\star=\arg\max_{\bar x}\ \frac{e^{-\alpha\bar x}}{D(\bar x)^2+\varrho_0^2},\qquad
D(\bar x)=\max\big(|\bar x-x_{\mathrm{lo}}|,|\bar x-x_{\mathrm{hi}}|\big),
\label{eq:robctr}
\end{equation}
with $\varrho_0^2$ the worst-case transverse term, which is quasi-concave in $\bar x$ and
balances the Chebyshev center of the cell against the attenuation gradient.
\end{proposition}
\begin{IEEEproof}
For a fixed placement, $\Theta$ as a function of user position is the product of a unimodal
focusing factor $G_s$ and a monotone pathloss factor $e^{-\alpha\bar x}/d^2$, hence
quasi-concave on the convex cell, so its minimum is attained on the boundary. If the axial
diameter does not exceed $W_{\mathrm{spatial}}$, then $G_s\ge1/2$ throughout $\mathcal U$ by
Theorem~\ref{thm:spatial}, so the focal-centered placement loses at most $3$~dB. Beyond the
focal spot the binding term is pathloss, and for fixed $\bar x$ the worst user maximizes
distance, giving the inner value $e^{-\alpha\bar x}/(D(\bar x)^2+\varrho_0^2)$. Maximizing over
$\bar x$ yields \eqref{eq:robctr}, whose logarithm $-\alpha\bar x-\ln(D(\bar x)^2+\varrho_0^2)$
is concave in $\bar x$ with $D$ convex and piecewise linear, so the maximizer trades the
Chebyshev center against a smaller $\bar x$.
\end{IEEEproof}
\noindent
For a cell much larger than the focal spot, multiple co-phased clusters tile $\mathcal U$, each
covering one focal spot. In all cases the feed delays remain closed-form, since they depend on
cluster geometry rather than on the user, and the intra-cluster orders remain minimal-step by
Proposition~\ref{prop:consec}, so robustness is carried entirely by the cluster center and the
number of tiling clusters.

The results above assemble into a solver that avoids any search over the oscillatory landscape.
Each cluster centroid is sited at its pathloss optimum (or the cell is tiled by focal spots,
Proposition~\ref{prop:robust}); the feasible step
$q^\star=\lceil\Delta_{\min}/\ell\rceil$ and the optimal orders $m_i=m_0+(i-1)q^\star$ are set
in closed form; the lattice positions follow from the monotone electrical path by
one-dimensional root finding; the feed delays are closed-form; and the sub-wavelength offsets
are either zero under Assumption~\ref{as:amp}, or refined by the SCA update \eqref{eq:scaupdate}
when the amplitude taper is non-negligible. The procedure is Algorithm~\ref{alg:structured},
with cost $\mathcal O(W I_{\mathrm{SCA}}(N_wK+N_w^{1.5}))$ when offsets are refined and
$\mathcal O(N\log\epsilon^{-1})$ when they vanish, in either case polynomial in $N$, $W$, $K$,
against the $\mathcal O((\neff D_x/\lambda_c)^N)$ local optima of $(\mathrm P)$ and the
$\mathcal O(N)$ delay elements of a conventional delay-phase array \cite{dai2021delay}.

\begin{algorithm}[t]
\caption{Structured Wideband PASS Placement}
\footnotesize
\label{alg:structured}
\begin{algorithmic}[1]
\Require user $(x_m,y_m)$ (or cell $\mathcal U$), $N$, $W$, $\Delta_{\min}$,
         subcarriers $\{f_k\}$, guide $(\neff,\ngr,\beta_2,\alpha_0,s)$
\Ensure  pinch positions $\{x_{w,i}\}$, feed delays $\{\tau_w\}$
\State partition the $N$ pinches into $W$ clusters of size $N_w=N/W$; site each
       centroid $\bar x_w$ at the pathloss optimum near the user (tile
       $\mathcal U$ if $\mathrm{diam}(\mathcal U)>\Wsp$,
       Prop.~\ref{prop:robust})
\For{each cluster $w=1,\dots,W$}
  \State $\cos\zeta_w\!\gets\!(x_m-\bar x_w)/d_w$;\ \
         $\ell_w\!\gets\!\lambda_c/(\neff-\cos\zeta_w)$
  \State $q^\star_w\gets\lceil\Delta_{\min}/\ell_w\rceil$
         \Comment{feasible step, \eqref{eq:qstar}}
  \State $m_i\gets -\tfrac{(N_w-1)q^\star_w}{2}+(i-1)q^\star_w$
         \Comment{optimal orders, Prop.~\ref{prop:consec}}
  \State $x^{(0)}_{w,i}\gets P^{-1}\!\big(P_0+m_i\lambda_c\big)$ by bisection
         \Comment{$P$ monotone, Thm.~\ref{thm:mip}}
  \State $\bar\tau_w\gets\frac1{N_w}\sum_i(d_{w,i}+\ngr x^{(0)}_{w,i})/c$
\EndFor
\State $\tau_w\gets \max_{w'}\bar\tau_{w'}-\bar\tau_w$
       \Comment{causal feed delays, Prop.~\ref{prop:ttd}}
\If{amplitude taper negligible (Assumption~\ref{as:amp})}
  \State $\delta_{w,i}\gets 0$
\Else
  \State refine $\{\delta_{w,i}\}$ by the SCA update \eqref{eq:scaupdate}
         to convergence
\EndIf
\State \Return $x_{w,i}=x^{(0)}_{w,i}+\delta_{w,i}$,\ \ $\{\tau_w\}$
\end{algorithmic}
\end{algorithm}

\subsection{Optimization Benchmarks}\label{subsec:cert}
Two baselines validate Algorithm~\ref{alg:structured}. The first is an exhaustive lattice
search that, made finite by Theorem~\ref{thm:mip}, \emph{certifies} global optimality. Since
every candidate optimum lies on the co-phasing lattice and the feed delays of
Proposition~\ref{prop:ttd} decouple the clusters, for a cluster of $N_w$ pinches one enumerates
all feasible order sets (strictly increasing integer vectors with adjacent gaps at least
$q^\star$) within a window of $M$ lattice cells and returns the maximizer of $\Theta$. This
visits $\binom{M}{N_w}$ candidates and finds the global optimum exactly, a certificate against
which any heuristic can be checked and one with no counterpart in a continuous-phase array. The
enumeration is tractable because wideband coherence forces the cluster to be small
(Remark~\ref{rem:aperture}).

The second baseline\label{subsec:pso} applies particle swarm optimization (PSO) directly to the
continuous, non-convex $(\mathrm P)$, representative of the metaheuristics usually entrusted
with oscillatory wideband-array objectives \cite{bereyhi2025mimopass}. Each particle encodes a
full placement $\mathbf p=[x_1,\dots,x_N,\tau_1,\dots,\tau_W]$, scored by the penalized fitness
\begin{equation}
\mathcal F(\mathbf p)=\Theta(\mathbf p)
-\lambda_{\mathrm{pen}}\!\sum_{i}\big[\Delta_{\min}-(x_{(i+1)}-x_{(i)})\big]_+^2,
\label{eq:psofit}
\end{equation}
with $x_{(i)}$ the sorted positions and $[\cdot]_+=\max(\cdot,0)$, updated by the canonical
velocity rule. PSO needs no structural knowledge but offers no guarantee and, as
Section~\ref{sec:num} shows, stalls well below the certified optimum because the
$\sim\!10^4$-per-coordinate local optima of $(\mathrm P)$ trap the swarm.
Table~\ref{tab:methods} contrasts the three strategies; the quantitative comparison is deferred
to Section~\ref{sec:num}.

\begin{table}[t]
\centering
\caption{Solution strategies for wideband placement.}
\label{tab:methods}
\renewcommand{\arraystretch}{1.25}
\begin{tabular}{p{2.15cm}p{2.7cm}p{2.4cm}}
\hline
Strategy & Complexity & Optimality \\
\hline
Proposed (Alg.~\ref{alg:structured}) &
$\mathcal O\!\big(W I_{\mathrm{SCA}}(N_wK+N_w^{1.5})\big)$ &
lattice-optimal (Prop.~\ref{prop:consec}) \\
Exhaustive lattice &
$\mathcal O\!\big(W\binom{M}{N_w}N_wK\big)$ &
global (certificate) \\
PSO on $(\mathrm P)$ &
$\mathcal O(I_{\mathrm{PSO}}\,n_{\mathrm{part}}\,NK)$ &
none \\
\hline
\end{tabular}
\end{table}

\section{Numerical Results}\label{sec:num}
This section validates the theoretical analysis through extensive Monte Carlo simulations. Throughout the numerical results, the analytical predictions are compared against independent simulations of the considered wideband channel model, demonstrating the accuracy of the derived expressions and the validity of the underlying theoretical approximations. Unless stated otherwise we adopt THz parameters drawn from the 6G and pinching-antenna literature. The system operates in the sub-THz 6G band at $f_c=300$~GHz over $B=10$~GHz~\cite{sarieddeen2021overview,enad2025theoretical,11006064,enad2026symbolerroranalysislinear}, on a low-loss dielectric guide with $\neff=1.4$, $\ngr=1.5$~\cite{rodrigues2023dielectric,ouyang2025array}, and GVD parameter $\beta_2=1$~ps$^2$/m referenced and swept toward larger near-cutoff values \cite{balanis2012advanced,yeh2008essence}. The user lies in a $10\times10$~m service region at array height $h=5$~m~\cite{tyrovolas2026performance}, with minimum pinch spacing $\Delta_{\min}=\lambda_c/2$ \cite{ouyang2025array}.
Ergodic quantities are averaged over $N_{\mathrm{mc}}=10^4$ user drops with
$x_m\sim\mathcal U[0,D_x]$, $y_m\sim\mathcal U[-D_y/2,D_y/2]$
\cite{tyrovolas2026performance}.

\begin{figure}[t]
\vspace{-5mm}
\centering
\includegraphics[width=0.9\columnwidth]{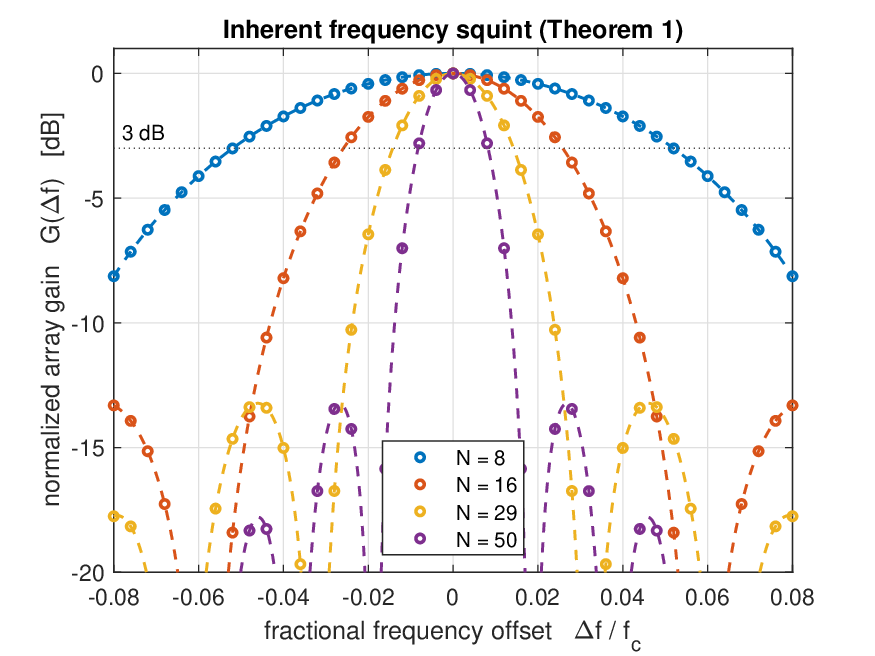}
\caption{Normalized array gain vs.\ fractional offset: full model (markers)
vs.\ Dirichlet closed form (dashed).}
\label{fig:dirichlet}
\end{figure}

\begin{figure}[t]
\centering
\begin{subfigure}[t]{0.49\columnwidth}
  \centering
  \includegraphics[width=\linewidth]{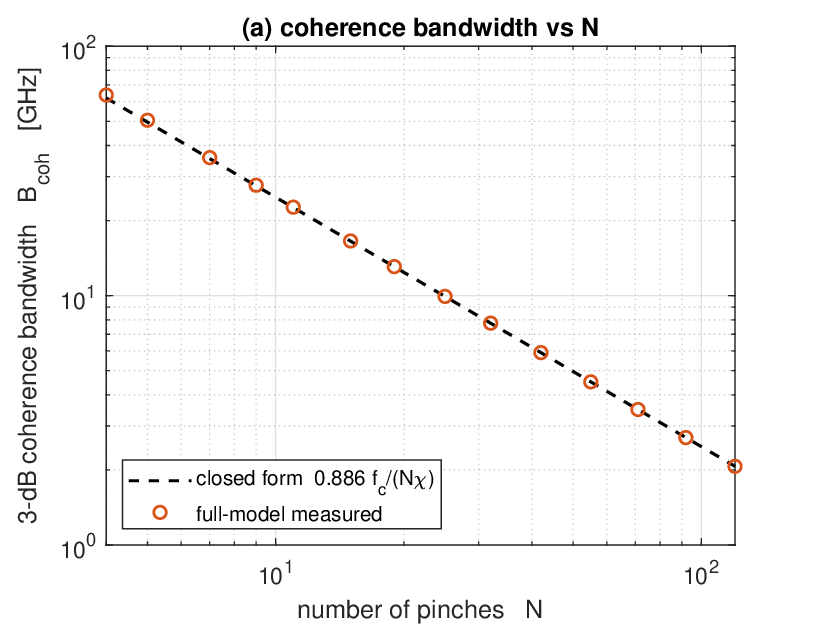}
  \caption{$\Bcoh$ vs.\ $N$.}
  \label{fig:bcoh_a}
\end{subfigure}
\hfill
\begin{subfigure}[t]{0.49\columnwidth}
  \centering
  \includegraphics[width=\linewidth]{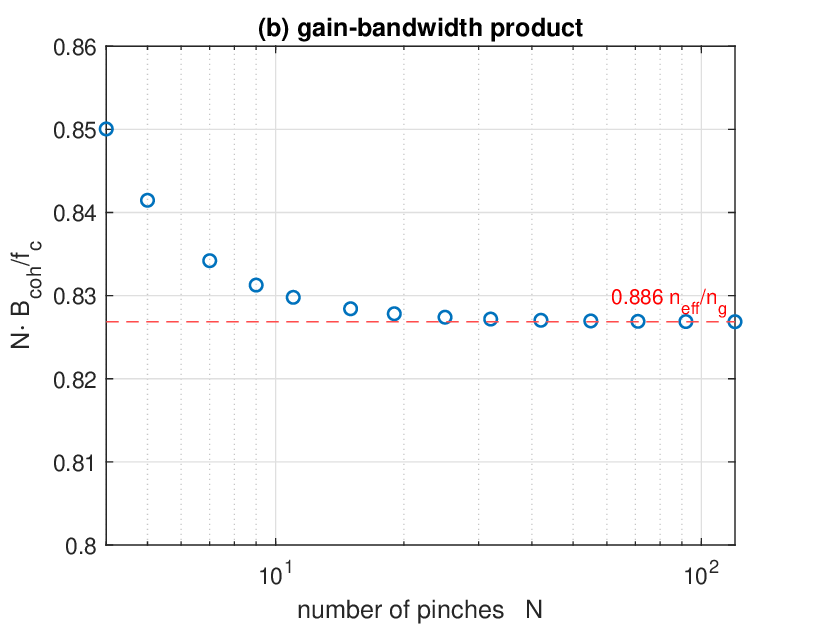}
  \caption{$N\,\Bcoh/f_c$ vs.\ $N$.}
  \label{fig:bcoh_b}
\end{subfigure}
\caption{Coherence bandwidth and its gain--bandwidth invariance.}
\label{fig:bcoh}
\vspace{-4mm}
\end{figure}

\paragraph{Inherent Squint and Coherence Bandwidth}\label{subsec:num_squint}
Fig.~\ref{fig:dirichlet} overlays the Dirichlet closed form \eqref{eq:dirichlet} on the full
model, which coincide to marker width for every $N\in\{8,16,29,50\}$, so the kernel is exact and
the array scans in $\Delta f/f_c$ rather than angle. Fig.~\ref{fig:bcoh} turns this into a design
law: panel~(a) confirms $\Bcoh\approx0.886f_c/(N\chi)$ and its $-1$ slope, so coherent combining
over $B$ needs $N\lesssim0.886f_c/(\chi B)$, with dispersion costing a factor $\neff/\ngr$ at
broadside; panel~(b) shows the product $N\cdot\Bcoh/f_c$ flat at $0.886\,\neff/\ngr$
\eqref{eq:gbw}, the $p=1$ case of Proposition~\ref{prop:law}, so gain and coherence bandwidth are
one operating point on a fixed hyperbola that only TTD correction leaves.

\begin{figure}[t]
\vspace{-5mm}
\centering
\begin{subfigure}[t]{0.49\columnwidth}
  \centering
  \includegraphics[width=\linewidth]{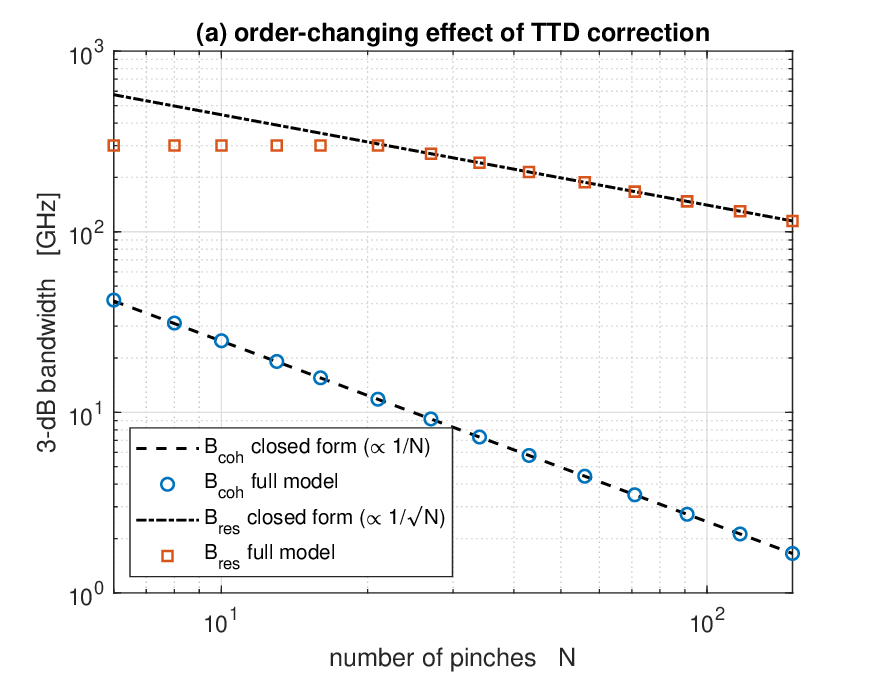}
  \caption{$\Bcoh\propto1/N$ vs.\ $\Bres\propto1/\sqrt N$.}
  \label{fig:floor_a}
\end{subfigure}
\hfill
\begin{subfigure}[t]{0.49\columnwidth}
  \centering
  \includegraphics[width=\linewidth]{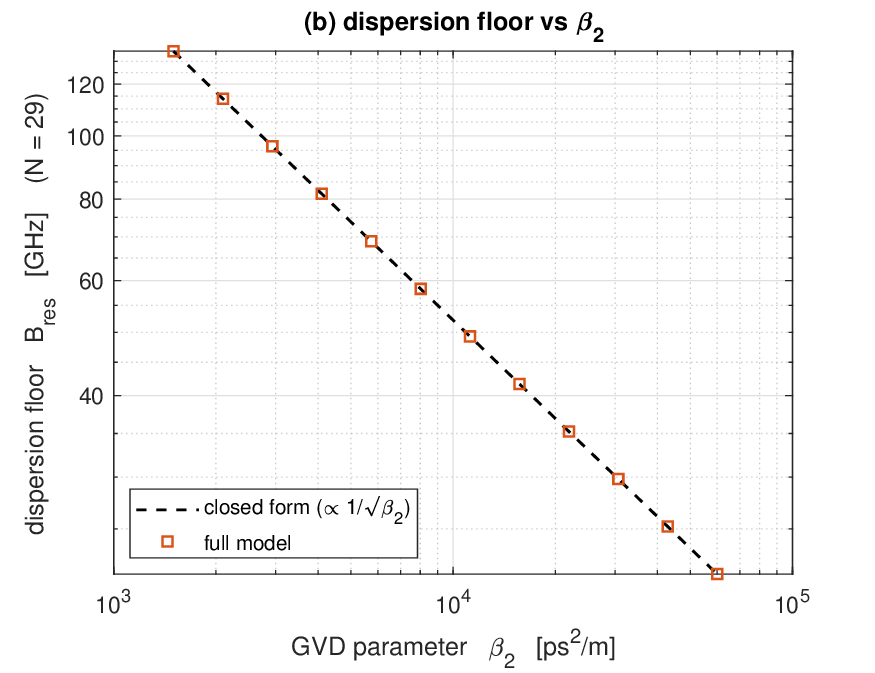}
  \caption{$\Bres\propto1/\sqrt{\beta_2}$, $N=29$.}
  \label{fig:floor_b}
\end{subfigure}
\caption{Dispersion floor and the order-changing effect of TTD correction.}
\label{fig:floor}
\end{figure}

\paragraph{The Irreducible Dispersion Floor}\label{subsec:num_floor}
Fig.~\ref{fig:floor}(a) places the uncorrected $\Bcoh$ and the post-TTD floor
$\Bres$ on log--log axes, with slopes $-1$ and $-\tfrac12$: TTD correction changes
the \emph{exponent} of the gain--bandwidth trade, not merely its constant, so the
product grows as $\sqrt N$. At the nominal near-lossless guide the floor sits in the
THz range (the crossover $\Delta f^\star=\ngr/(\pi\beta_2 c)$ exceeds $\Bcoh$), so
an enlarged near-cutoff $\beta_2$ is used to bring it onto the axes; dispersion is
then not the operative constraint, and after TTD the pinch count is spacing-limited,
not dispersion-limited. Fig.~\ref{fig:floor}(b) confirms $\Bres\propto1/\sqrt{\beta_2}$
over a decade.

\begin{figure}[t]
\centering
\begin{subfigure}[t]{0.49\columnwidth}
  \centering
  \includegraphics[width=\linewidth]{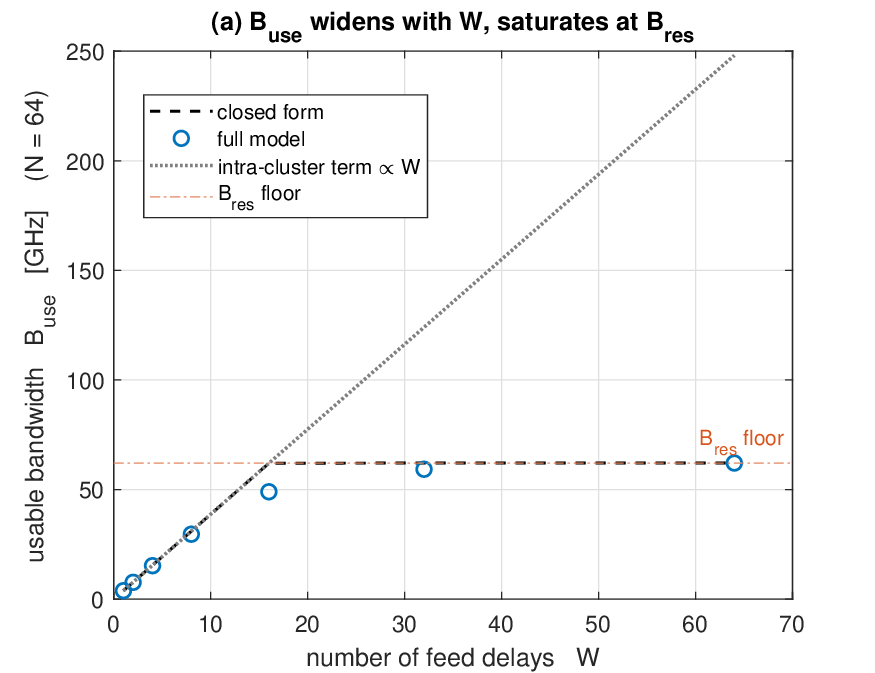}
  \caption{$\Buse$ vs.\ $W$, $N=64$.}
  \label{fig:wmin_a}
\end{subfigure}
\hfill
\begin{subfigure}[t]{0.49\columnwidth}
  \centering
  \includegraphics[width=\linewidth]{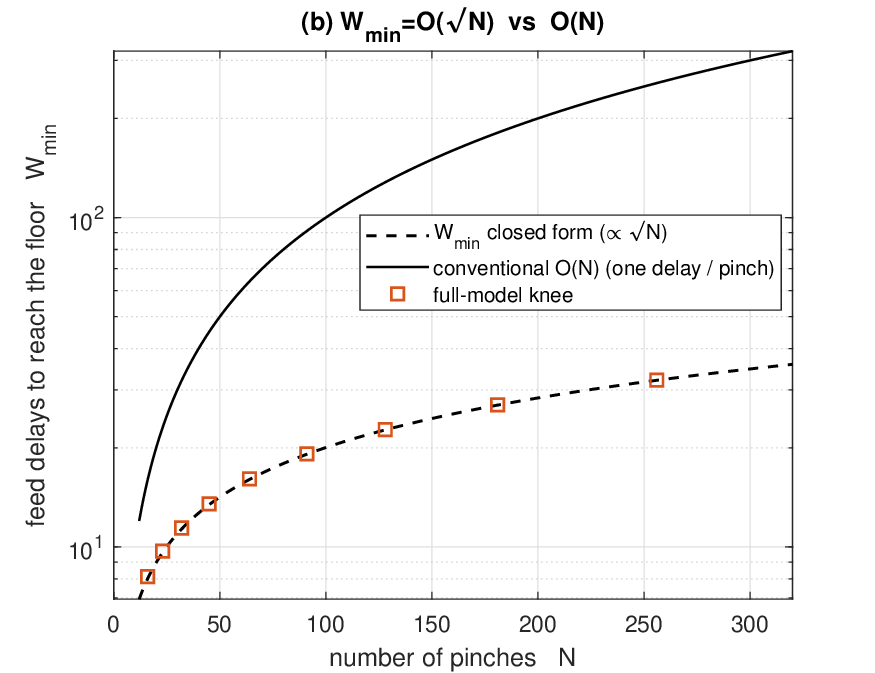}
  \caption{$W_{\min}=\mathcal O(\sqrt N)$ vs.\ $\mathcal O(N)$.}
  \label{fig:wmin_b}
\end{subfigure}
\caption{Delay-phase mitigation: usable band and the delays needed to reach it.}
\label{fig:wmin}
\vspace{-4mm}
\end{figure}

\paragraph{Delay-Phase Mitigation with Few Delays}\label{subsec:num_ttd}
Fig.~\ref{fig:wmin}(a) tracks the usable bandwidth $\Buse$ as feed delays are
added. Two regimes meet at a sharp knee. For small $W$, $\Buse$ grows linearly:
each delay halves the sub-cluster size and, since the size-$N/W$ squint scales as
$0.886f_c W/N\cdot\neff/\ngr$, doubles the coherent band. Past the knee $\Buse$
saturates at the dispersion floor $\Bres$, where the full-aperture GVD curvature
takes over and further delays are wasted; the full-model markers follow the
$\min\{\cdot,\cdot\}$ envelope of \eqref{eq:buse}. The knee is the point of
diminishing returns on delay hardware. Fig.~\ref{fig:wmin}(b) shows the delays
needed to reach the floor, $W_{\min}$, following the $\mathcal O(\sqrt N)$ law
\eqref{eq:wmin} against the $\mathcal O(N)$ line of a conventional delay-phase
array: the delayed fraction $W_{\min}/N=\mathcal O(1/\sqrt N)$ vanishes with array
size. Pinch positioning provides per-element fine alignment, leaving the feed delays responsible only for the $\mathcal{O}(\sqrt{N})$ inter-cluster offsets. As in Fig.~\ref{fig:floor}, an enlarged near-cutoff $\beta_2$ brings the floor into the plotted range; the $W/N$ rise and $\sqrt{N}$ scaling are independent of $\beta_2$, which determines only the floor height.

\begin{figure}[t]
\vspace{-5mm}
\centering
\begin{subfigure}[t]{0.49\columnwidth}
  \centering
  \includegraphics[width=\linewidth]{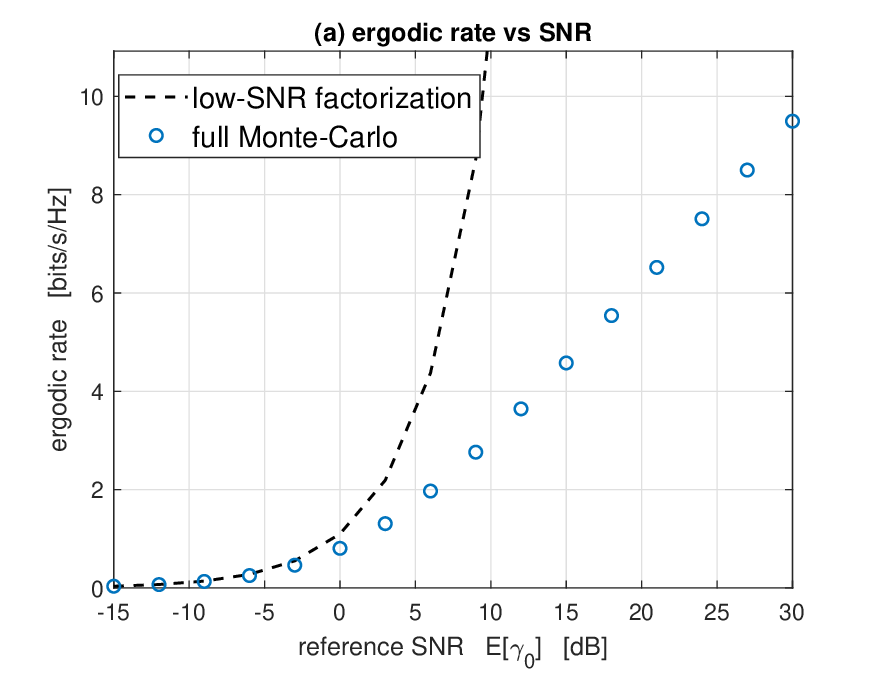}
  \caption{Rate vs.\ $\mathbb E[\gamma_0]$.}
  \label{fig:rate_a}
\end{subfigure}
\hfill
\begin{subfigure}[t]{0.49\columnwidth}
  \centering
  \includegraphics[width=\linewidth]{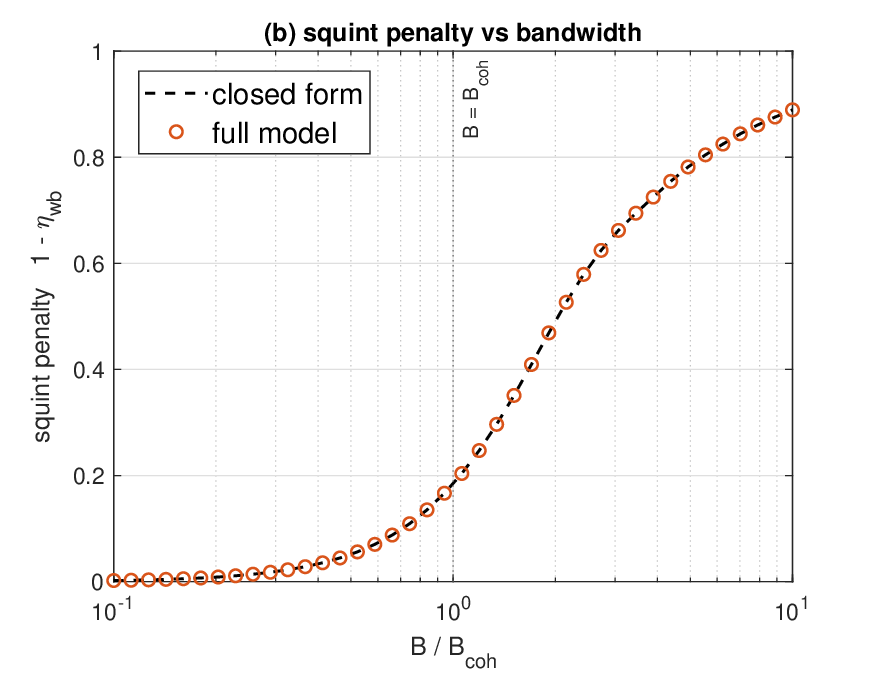}
  \caption{Penalty $1-\etawb$ vs.\ $B/\Bcoh$.}
  \label{fig:rate_b}
\end{subfigure}
\caption{Wideband ergodic rate and the squint penalty.}
\vspace{-3mm}
\label{fig:rate}
\end{figure}

\paragraph{Wideband Ergodic Rate and Molecular Absorption}\label{subsec:num_rate}
Fig.~\ref{fig:rate}(a) validates the structural claim of Theorem~\ref{thm:lowsnr}:
the user-position statistics and the wideband squint separate. The full Monte-Carlo
matches the factorization $\bar R\approx(\etawb/\ln2)\,\mathbb E[\gamma_0]$ through
the low-to-moderate SNR range, departing only above $0$~dB where the logarithm
compresses. Since THz links are power- and distance-limited, this is exactly
the operative regime: the wideband rate collapses to a product of a geometry-only
functional $\mathbb E[\gamma_0]$ \eqref{eq:Egamma0} and a deterministic squint
scalar $\etawb$, so array sizing and site planning decouple. Fig.~\ref{fig:rate}(b)
plots the penalty $1-\etawb$ against $B/\Bcoh$: small and flat below
$B\!\approx\!\Bcoh$, rising steadily beyond it as subcarriers spill outside the main
lobe. This identifies $B\approx\Bcoh$ as the economical operating point and
motivates the pinch-count optimization of Section~\ref{subsec:num_Nstar}.

\begin{figure}[t]
\vspace{-1mm}
\centering
\begin{subfigure}[t]{0.49\columnwidth}
  \centering
  \includegraphics[width=\linewidth]{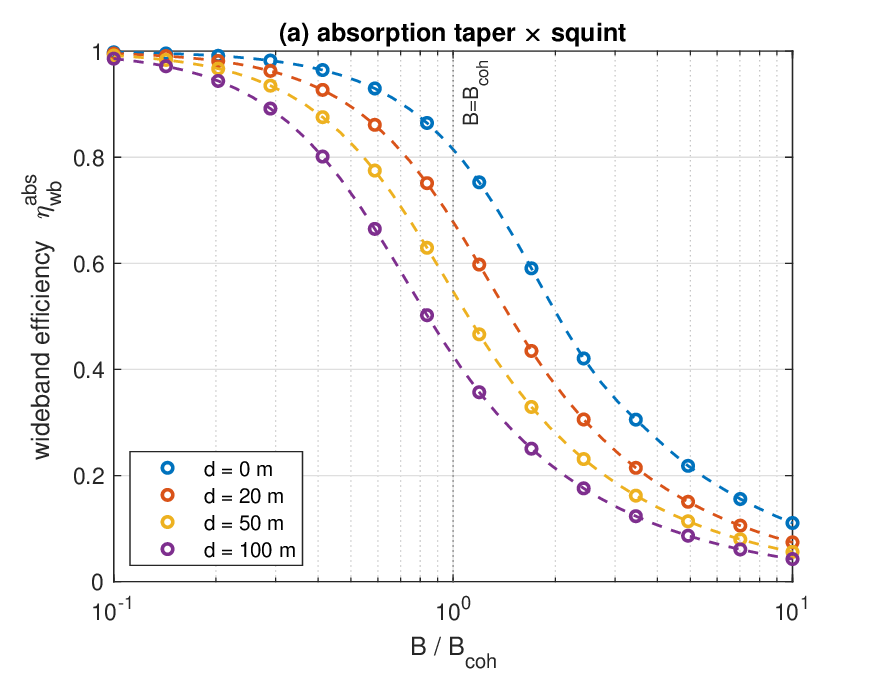}
  \caption{$\etawb^{\mathrm{abs}}$ vs.\ $B/\Bcoh$.}
  \label{fig:absorption_a}
\end{subfigure}
\hfill
\begin{subfigure}[t]{0.49\columnwidth}
  \centering
  \includegraphics[width=\linewidth]{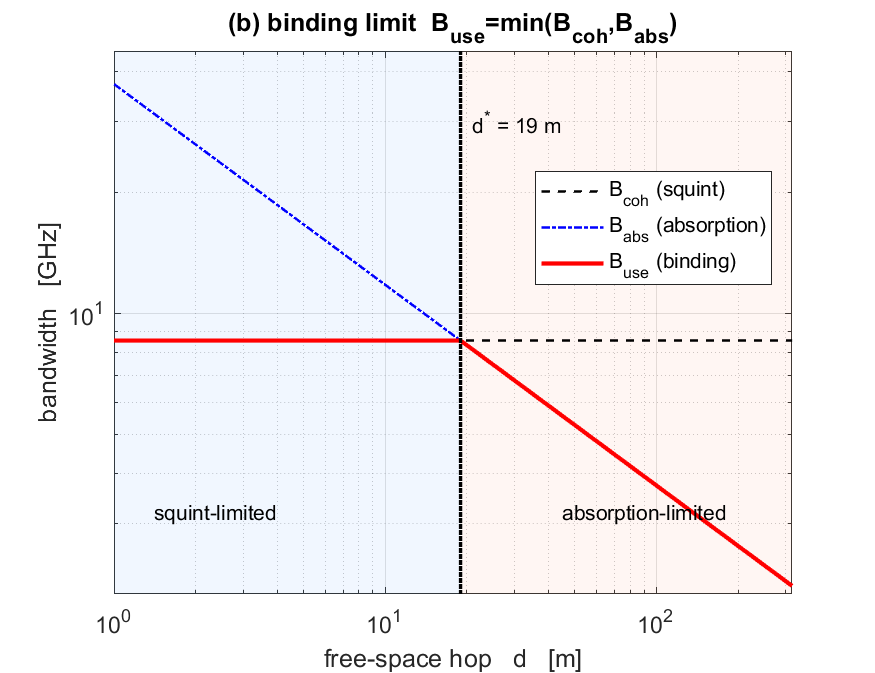}
  \caption{$\Buse=\min\{\Bcoh,\Babs\}$ vs.\ $d$.}
  \label{fig:absorption_b}
\end{subfigure}
\caption{Molecular absorption as a deterministic taper multiplying the squint.}
\label{fig:absorption}
\vspace{-5mm}
\end{figure}

Fig.~\ref{fig:absorption} adds molecular absorption. Panel~(a) shows
$\etawb^{\mathrm{abs}}$ for several hop distances: as $d$ grows the taper
$e^{-[\kappa(f)-\kappa(f_c)]d}$ pulls efficiency down faster in $B$ but never
reshapes the squint roll-off, confirming that absorption multiplies the squint
kernel as an independent factor. Panel~(b) makes the trade explicit through
$\Buse=\min\{\Bcoh,\Babs\}$: $\Bcoh$ is $d$-independent while
$\Babs=2.355/\sqrt{\kappa_2 d}$ falls with distance, and their crossover
$d^\star$ \eqref{eq:dstar} splits the deployment into a squint-limited regime
($d<d^\star$) and an absorption-limited one ($d>d^\star$). Over the short
LOS hops PASS is built to create, the operating point is squint-limited,
so dispersion-driven squint is the wideband bottleneck; absorption dominates only
on long hops or near a window edge, where $\kappa_2$ is large and $d^\star$
contracts toward the metre scale shown.

\begin{figure}[t]
\vspace{-5mm}
\centering
\begin{subfigure}[t]{0.49\columnwidth}
  \centering
  \includegraphics[width=\linewidth]{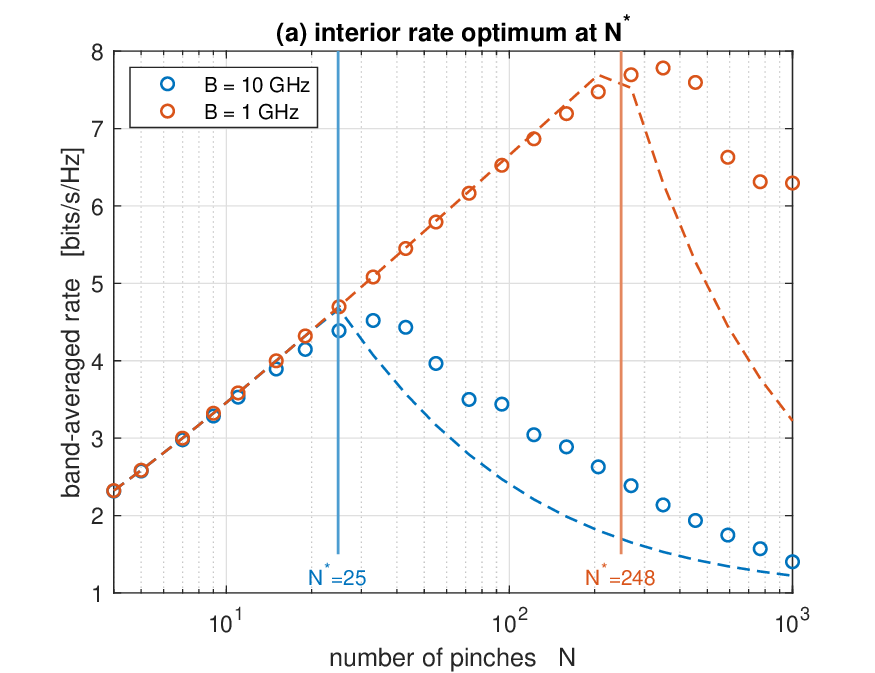}
  \caption{Rate vs.\ $N$, $B\in\{1,10\}$~GHz.}
  \label{fig:nstar_a}
\end{subfigure}
\hfill
\begin{subfigure}[t]{0.49\columnwidth}
  \centering
  \includegraphics[width=\linewidth]{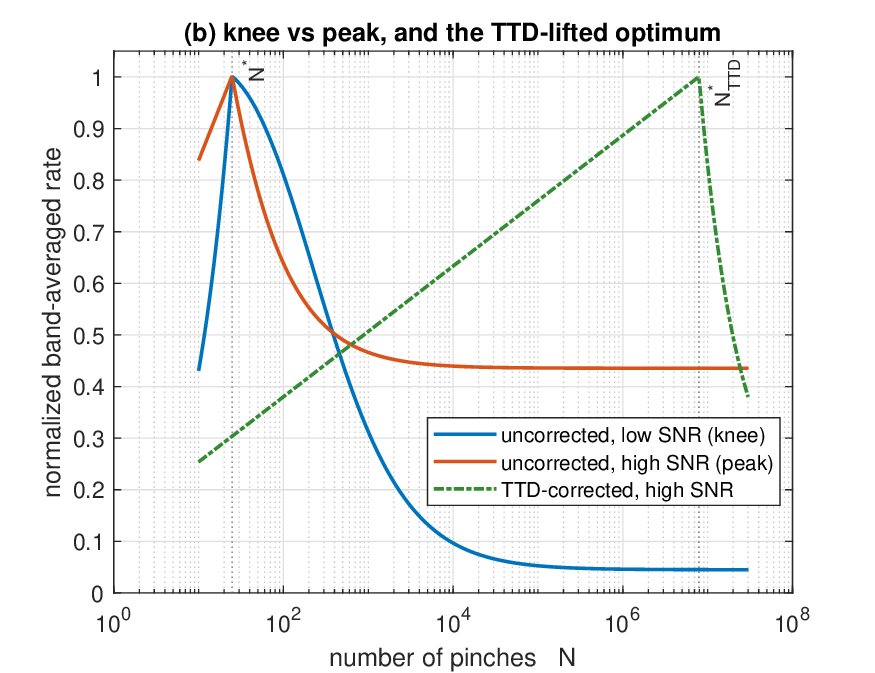}
  \caption{Knee vs.\ peak; the TTD-lifted $N^\star_{\mathrm{TTD}}$.}
  \label{fig:nstar_b}
\end{subfigure}
\caption{Rate-optimal pinch count and its TTD lift
(Proposition~\ref{prop:Nstar}, Corollary~\ref{cor:Nttd}).}
\label{fig:Nstar}
\end{figure}

\paragraph{Rate-Optimal Pinch Count}\label{subsec:num_Nstar}
Because the coherent band-centre SNR grows as $Ng$ while $\Bcoh\propto1/N$, the
band-averaged rate has an interior optimum. Fig.~\ref{fig:Nstar}(a) shows the rate
rising while the band stays coherent, then flattening into a knee at the closed-form
$N^\star=0.886f_c/(\chi B)$ \eqref{eq:Nstar}, where $\Bcoh(N^\star)=B$, for both
$B=1$ and $10$~GHz, confirming Proposition~\ref{prop:Nstar}: size the array so
coherence bandwidth just meets signal bandwidth. Fig.~\ref{fig:Nstar}(b)
distinguishes two regimes. At low SNR the curve is a knee, over-provisioning costs
only second order in the per-pinch SNR, so a power-limited THz array is
forgiving. At high SNR it is a genuine peak, where over-provisioning wastes array
gain on incoherent subcarriers; such links must size precisely at $N^\star$. The
panel also shows that replacing $\Bcoh$ by $\Bres\propto1/\sqrt N$ moves the knee to
$N^\star_{\mathrm{TTD}}=0.564(\neff-\cos\zeta)f_c/(\beta_2 c B^2)$, orders of
magnitude higher: after TTD the useful pinch count is limited by the spacing and
aperture feasibility of Section~\ref{subsec:num_spatial}, not dispersion.

\begin{figure}[t]
\centering
\begin{subfigure}[t]{0.49\columnwidth}
  \centering
  \includegraphics[width=\linewidth]{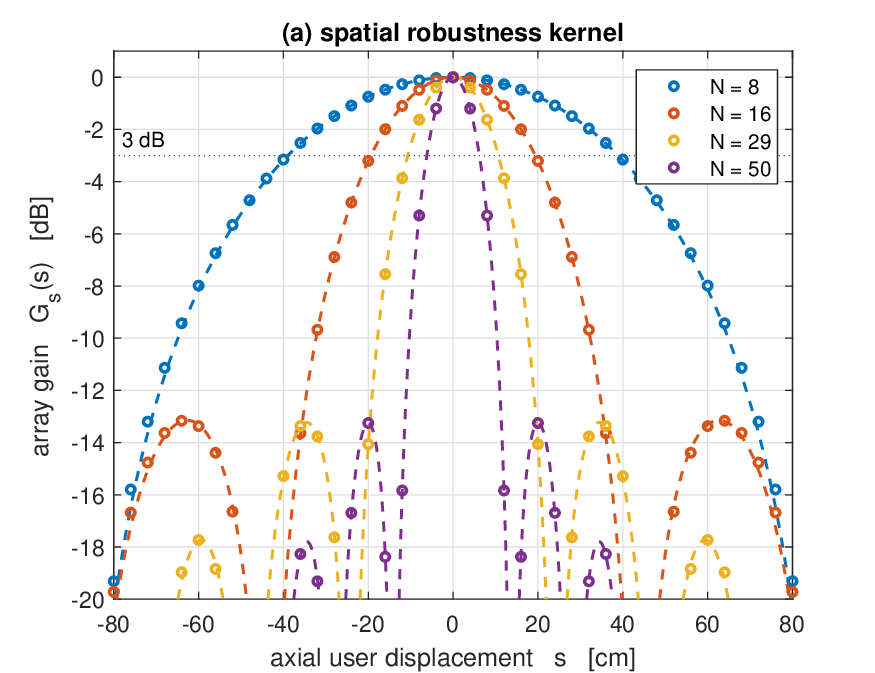}
  \caption{Gain vs.\ displacement $s$.}
  \label{fig:spatial_a}
\end{subfigure}
\hfill
\begin{subfigure}[t]{0.49\columnwidth}
  \centering
  \includegraphics[width=\linewidth]{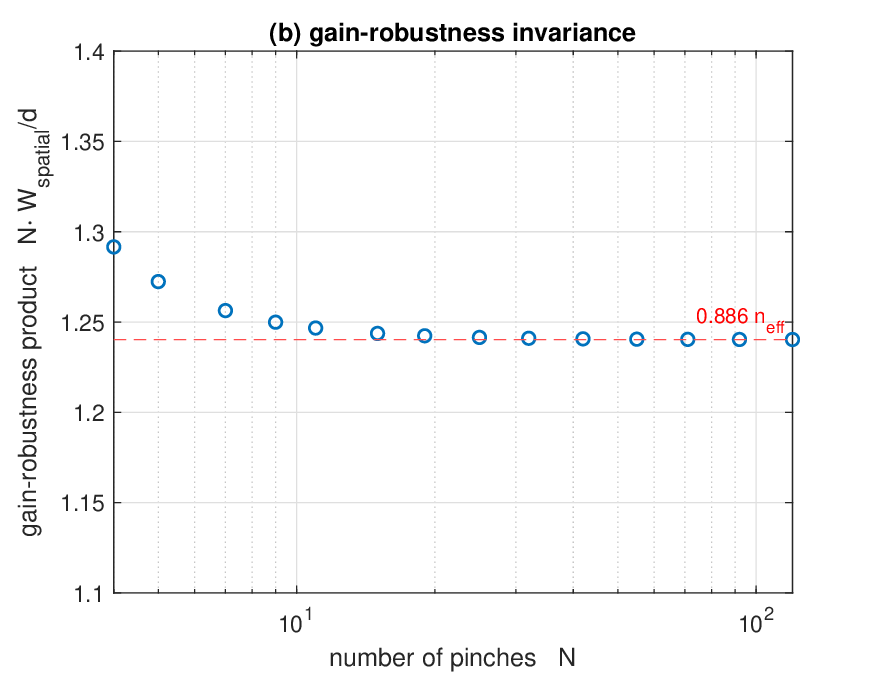}
  \caption{$N\,\Wsp/d = 0.886\,\neff$.}
  \label{fig:spatial_b}
\end{subfigure}
\caption{Spatial robustness and its gain--robustness invariance.}
\label{fig:spatial}
\vspace{-2mm}
\end{figure}

\begin{figure}[t]
\centering
\begin{subfigure}[t]{0.49\columnwidth}
  \centering
  \includegraphics[width=\linewidth]{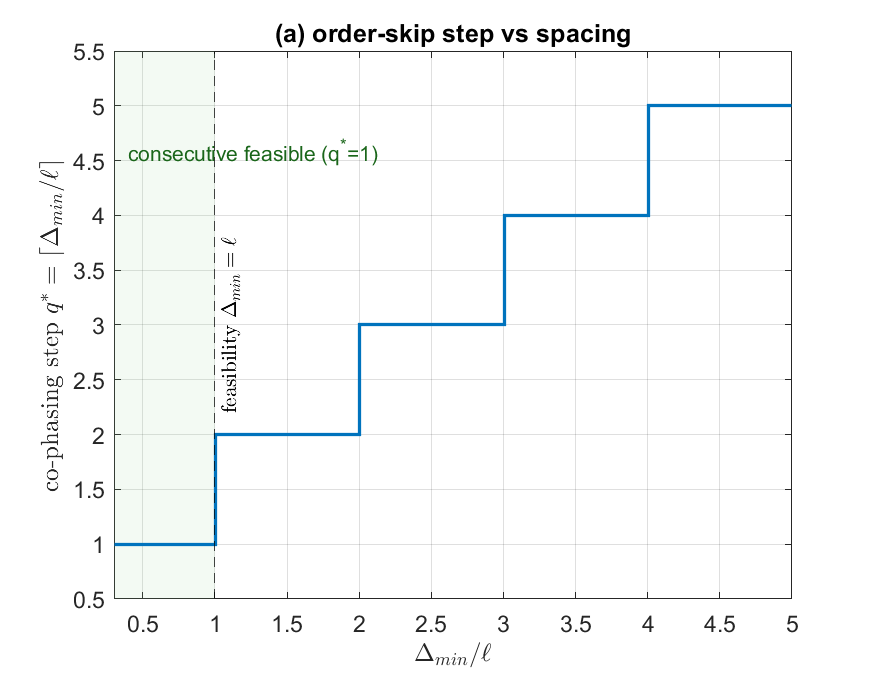}
  \caption{$q^\star$ vs.\ $\Delta_{\min}/\ell$.}
  \label{fig:qstar_a}
\end{subfigure}
\hfill
\begin{subfigure}[t]{0.49\columnwidth}
  \centering
  \includegraphics[width=\linewidth]{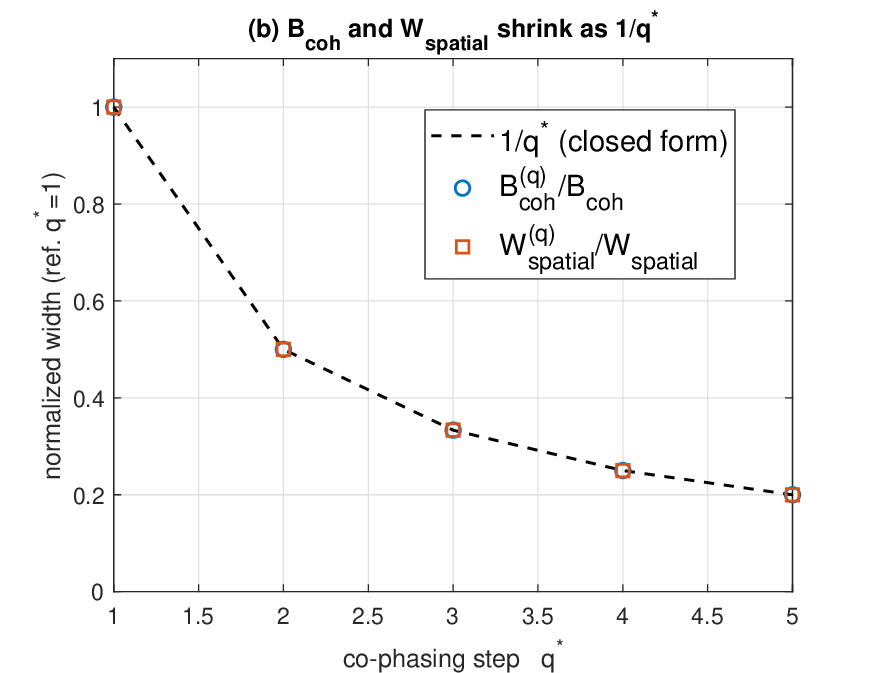}
  \caption{$\Bcoh,\Wsp$ shrink as $1/q^\star$.}
  \label{fig:qstar_b}
\end{subfigure}
\caption{Order skipping under coupling-limited spacing.}
\label{fig:qstar}
\end{figure}

\paragraph{Spatial Robustness and Order Skipping}\label{subsec:num_spatial}
Fig.~\ref{fig:spatial}(a) plots the carrier-frequency gain as the user is displaced
axially from the co-phasing position. The full-model markers fall on the spatial
Dirichlet kernel of Theorem~\ref{thm:spatial} for every $N$, the exact counterpart
of the frequency kernel with $s/(\neff d)$ in place of $\Delta f/f_c$: spatial
robustness and coherence bandwidth are one phenomenon in conjugate variables. The
main-lobe width $W_{\mathrm{spatial}}=0.886\,\neff d/N$ is of order decimetres, so a
cluster tolerates position errors of that scale within $3$~dB without tracking, and
larger cells are tiled by several co-phased clusters. Fig.~\ref{fig:spatial}(b)
shows the gain--robustness product $N\cdot W_{\mathrm{spatial}}/d$ flat at
$0.886\,\neff$, the mirror of the gain--bandwidth invariance and the same $p=1$
case of Proposition~\ref{prop:law}, so gain, coherence bandwidth, and position
robustness are three readings of one design choice.
Fig.~\ref{fig:qstar} treats the coupling constraint. The co-phasing cell
$\ell=\lambda_c/(\neff-\cos\zeta)$ is sub-wavelength ($\approx0.71\lambda_c$ at
broadside), so when $\Delta_{\min}>\ell$ the array must skip orders by
$q^\star=\lceil\Delta_{\min}/\ell\rceil$. Fig.~\ref{fig:qstar}(a) shows this step as
a staircase with the feasibility boundary at $\Delta_{\min}=\ell$ separating the
consecutive regime ($q^\star=1$, where
Sections~\ref{subsec:num_squint}--\ref{subsec:num_Nstar} hold verbatim) from the
order-skipped regime. Fig.~\ref{fig:qstar}(b) shows both $\Bcoh$ and
$W_{\mathrm{spatial}}$ contracting as $1/q^\star$, the full-model points tracking
the law across $q^\star=1,\dots,5$. The mechanism is a single aperture effect, skipping orders stretches the cluster to $q^\star L$, so element coupling, by setting $\Delta_{\min}$, imposes a minimum aperture that debits both the usable bandwidth and the robustness budget.

\begin{figure}[t]
\centering
\begin{subfigure}[t]{0.49\columnwidth}
  \centering
  \includegraphics[width=\linewidth]{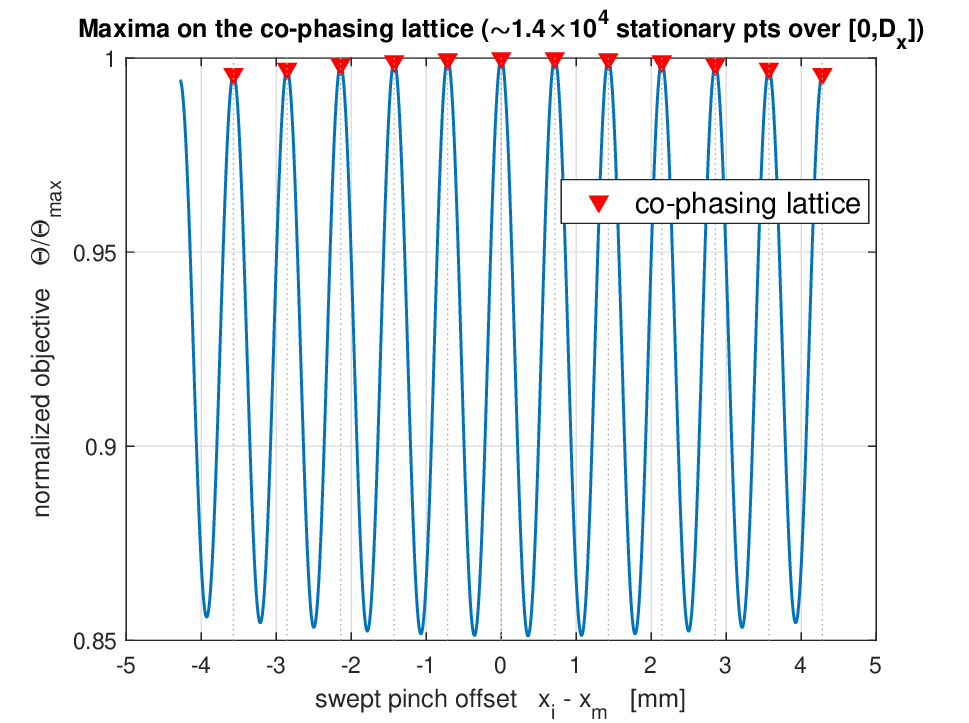}
  \caption{Non-convex landscape; maxima on the lattice.}
  \label{fig:landscape}
\end{subfigure}
\hfill
\begin{subfigure}[t]{0.49\columnwidth}
  \centering
  \includegraphics[width=\linewidth]{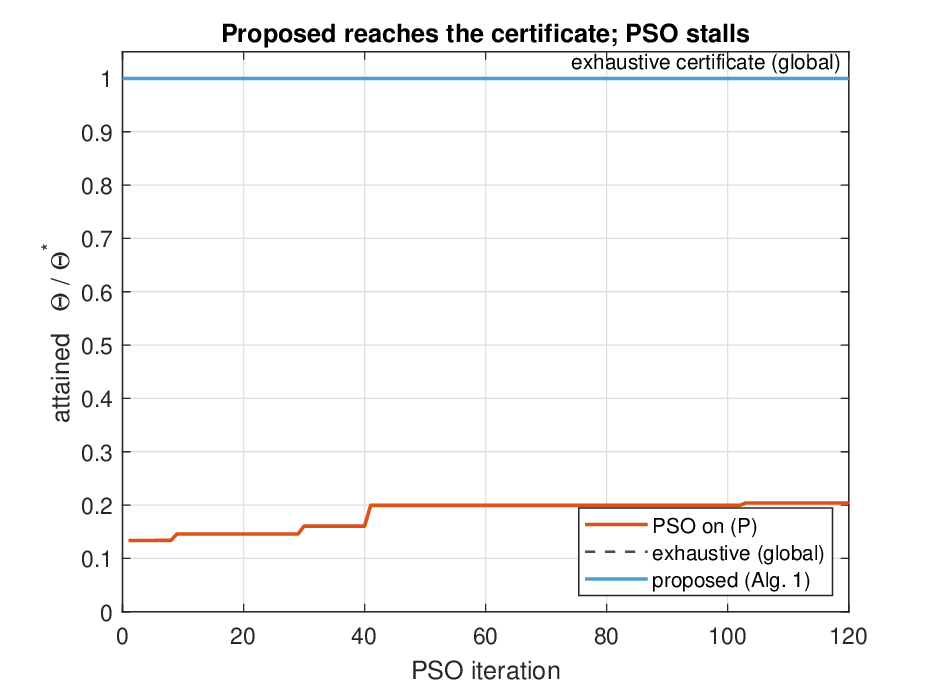}
  \caption{Attained $\Theta$ vs.\ budget.}
  \label{fig:optgap}
\end{subfigure}
\caption{Global-Optimality of the Proposed Placement Algorithm}
\label{fig:placement}
\end{figure}

\paragraph{Placement Optimization: Proposed vs.\ Exhaustive vs.\ PSO}
\label{subsec:num_placement}
Fig.~\ref{fig:placement}(a) shows why \eqref{eq:P} resists generic optimization and why its
integer structure is the way out. Sweeping one pinch across the region, $\Theta$ oscillates
$\sim\neff D_x/\lambda_c\approx1.4\times10^4$ times, giving as many stationary points per
coordinate and $\mathcal O((\neff D_x/\lambda_c)^N)$ joint local optima. Yet the maxima land
exactly on the co-phasing lattice points $x^{(0)}(m)$ of \eqref{eq:coarsefine}, so the
continuous search reduces without approximation to a choice of integer orders
(Theorem~\ref{thm:mip}), and one should snap positions to the lattice and optimize the orders.

Fig.~\ref{fig:placement}(b) compares the three strategies. The proposed algorithm places pinches
at minimal-step consecutive orders in closed form (Proposition~\ref{prop:consec}) and attains an
objective indistinguishable from the exhaustive certificate, so its optimality gap is
numerically zero ($0.00$~dB) at sub-millisecond cost for every $N$. PSO on the raw continuous
problem stalls below the certificate, trapped by the per-coordinate optima, and its gap widens
with array size from $-7.4$~dB at $N=8$ to $-10.8$~dB at $N=16$ and $-12.7$~dB at $N=29$.
Meanwhile the exhaustive runtime climbs steeply, from $9$~ms at $N=8$ to $0.6$~s at $N=29$,
whereas the proposed runtime stays near $0.2$~ms throughout. The structured method is thus the
only one that is both optimal and scalable.

\section{Conclusion}\label{sec:conclusion}
In this paper, we developed the first wideband, dispersion-aware theory of pinching-antenna systems, and showed that the pinching mechanism carries an inherent limitation. Because single-waveguide co-phasing equalizes path length only modulo a wavelength, a pinching array is intrinsically beam-squint limited, with a Dirichlet-kernel response, a closed-form coherence bandwidth, and an irreducible dispersion floor that even TTD correction cannot remove. A unifying law ties the behavior together, in which array gain trades against coherence bandwidth and against spatial robustness through one common invariant that TTD correction breaks, reaching the floor with far fewer delay elements than a conventional array. The law is actionable, yielding a rate-optimal pinch count where coherence meets signal bandwidth, a molecular absorption that multiplies the squint as a deterministic taper, and a placement problem that the integer structure of co-phasing renders an exactly solvable mixed-integer program with a provably optimal, polynomial-time solution.
The analysis is foundational, isolating the dispersion--squint mechanism through a single-user LOS link. The multi-user case, where per-user co-phasing lattices conflict, is the natural next step, alongside multipath channels, hardware impairments, and experimental validation.

\bibliographystyle{IEEEtran}
\bibliography{references}

@ARTICLE{rappaport2019wireless,
  author={Rappaport, Theodore S. and others},
  journal={IEEE Access}, 
  title={Wireless Communications and Applications Above 100 {GHz}: Opportunities and Challenges for {6G} and Beyond}, 
  year={2019},
  volume={7},
  number={},
  pages={78729-78757},
  doi={10.1109/ACCESS.2019.2921522}}

@ARTICLE{sarieddeen2021overview,
  author  = {Sarieddeen, Hadi and Alouini, Mohamed-Slim and Al-Naffouri, Tareq Y.},
  title   = {An Overview of Signal Processing Techniques for Terahertz Communications},
  journal = {Proc. IEEE},
  volume  = {109}, number = {10}, pages = {1628--1665}, year = {2021}
}

@ARTICLE{tarboush2021teramimo,
  author  = {Tarboush, Simon and others},
  title   = {{TeraMIMO}: A Channel Simulator for Wideband Ultra-Massive {MIMO} Terahertz Communications},
  journal = {IEEE Trans. Veh. Technol.},
  volume  = {70}, number = {12}, pages = {12325--12341}, year = {2021}
}

@ARTICLE{direnzo2020smart,
  author={Di Renzo, Marco and others},
  journal={IEEE J. Sel. Areas Commun.}, 
  title={Smart Radio Environments Empowered by {Reconfigurable Intelligent Surfaces}: How It Works, State of Research, and The Road Ahead}, 
  year={2020},
  volume={38},
  number={11},
  pages={2450-2525},
  doi={10.1109/JSAC.2020.3007211}}

@ARTICLE{huang2020holographic,
  author={Huang, Chongwen and others},
  journal={IEEE Wireless Commun.}, 
  title={Holographic {MIMO} Surfaces for {6G} Wireless Networks: Opportunities, Challenges, and Trends}, 
  year={2020},
  volume={27},
  number={5},
  pages={118-125},
  doi={10.1109/MWC.001.1900534}}

@ARTICLE{ma2024mimo,
  author={Ma, Wenyan and Zhu, Lipeng and Zhang, Rui},
  journal={IEEE Trans. Wireless Commun.}, 
  title={{MIMO} Capacity Characterization for Movable Antenna Systems}, 
  year={2024},
  volume={23},
  number={4},
  pages={3392-3407},
  doi={10.1109/TWC.2023.3307696}}

@ARTICLE{liu2025pass,
  author={Liu, Yuanwei and others},
  journal={IEEE Commun. Mag.}, 
  title={Pinching-Antenna Systems: Architecture Designs, Opportunities, and Outlook}, 
  year={2026},
  volume={64},
  number={1},
  pages={190-196},
  doi={10.1109/MCOM.001.2500037}}

@ARTICLE{ding2024flexible,
  author={Ding, Zhiguo and others},
  journal={IEEE Trans. Commun.}, 
  title={Flexible-Antenna Systems: A {Pinching-Antenna} Perspective}, 
  year={2025},
  volume={73},
  number={10},
  pages={9236-9253},
  doi={10.1109/TCOMM.2025.3555866}}

@ARTICLE{tyrovolas2026performance,
  author={Tyrovolas, Dimitrios and others},
  journal={IEEE Trans. Cogn. Commun. Netw.}, 
  title={Performance Analysis of {Pinching-Antenna} Systems}, 
  year={2026},
  volume={12},
  number={},
  pages={590-601},
  doi={10.1109/TCCN.2025.3564470}}

@ARTICLE{tegos2025minimum,
  author={Tegos, Sotiris A. and others},
  journal={IEEE Wireless Commun. Lett.}, 
  title={Minimum Data Rate Maximization for Uplink {Pinching-Antenna} Systems}, 
  year={2025},
  volume={14},
  number={5},
  pages={1516-1520},
  doi={10.1109/LWC.2025.3547956}}

@ARTICLE{wang2024noma,
  author={Wang, Kaidi and Ding, Zhiguo and Schober, Robert},
  journal={IEEE Wireless Commun. Lett.}, 
  title={Antenna Activation for {NOMA} Assisted {Pinching-Antenna} Systems}, 
  year={2025},
  volume={14},
  number={5},
  pages={1526-1530},
  doi={10.1109/LWC.2025.3548280}}

@ARTICLE{ouyang2025array,
  author={Ouyang, Chongjun and others},
  journal={IEEE Commun. Lett.}, 
  title={Array Gain for Pinching-Antenna Systems ({PASS})}, 
  year={2025},
  volume={29},
  number={6},
  pages={1471-1475},
  doi={10.1109/LCOMM.2025.3566299}}

@ARTICLE{bereyhi2025mimopass,
  author={Bereyhi, Ali and others},
  journal={IEEE Trans. Commun.}, 
  title={{MIMO-PASS}: Uplink and Downlink Transmission via {MIMO} {Pinching-Antenna} Systems}, 
  year={2026},
  volume={74},
  number={},
  pages={5701-5716},
  doi={10.1109/TCOMM.2026.3668156}}

@ARTICLE{ouyang2025capacity,
  author={Ouyang, Chongjun and others},
  journal={IEEE Trans. Wireless Commun.}, 
  title={Capacity Characterization of {Pinching-Antenna} Systems}, 
  year={2026},
  volume={25},
  number={},
  pages={10387-10404},
  doi={10.1109/TWC.2026.3653191}}

@misc{dofrobust2026,
      title={Design of {DoF}-Robust {MIMO} via Generalized Pinching Antennas}, 
      author={Zhiguo Ding and others},
      year={2026},
      eprint={2610.36670},
      archivePrefix={arXiv},
      primaryClass={cs.IT},
      url={https://arxiv.org/abs/2610.36670}, 
}

@ARTICLE{dai2021delay,
  author={Dai, Linglong and others},
  journal={IEEE Trans. Wireless Commun.}, 
  title={Delay-Phase Precoding for Wideband {THz} Massive {MIMO}}, 
  year={2022},
  volume={21},
  number={9},
  pages={7271-7286},
  doi={10.1109/TWC.2022.3157315}}

@misc{sixdma2026,
      title={{Terahertz} Beam Squint Mitigation via Six-Dimensional {Movable Antennas}}, 
      author={Yike Xie and others},
      year={2026},
      eprint={2603.23859},
      archivePrefix={arXiv},
      primaryClass={eess.SP},
      url={https://arxiv.org/abs/2603.23859}, 
}

@book{balanis2012advanced,
  title={Advanced Engineering Electromagnetics},
  author={Balanis, Constantine A},
  year={2024},
  publisher={John Wiley \& Sons, Hoboken, NJ}
}

@book{yeh2008essence,
  title     = {The Essence of Dielectric Waveguides},
  author    = {Yeh, Cavour and Shimabukuro, Fred I.},
  publisher = {Springer}, 
  year = {2008}
}

@article{razaviyayn2013unified,
author = {Razaviyayn, Meisam and Hong, Mingyi and Luo, Zhi-Quan},
title = {A Unified Convergence Analysis of Block Successive Minimization Methods for Nonsmooth Optimization},
journal = {SIAM J. Optim.},
volume = {23},
number = {2},
pages = {1126-1153},
year = {2013},
doi = {10.1137/120891009}
}

@book{boyd2004convex,
  author    = {Boyd, Stephen and others},
  title     = {Convex Optimization},
  publisher = {Cambridge Univ. Press}, year = {2004}
}

@article{rodrigues2023dielectric,
  author  = {J. R. Rodrigues and others},
  title   = {All-dielectric scale invariant waveguide},
  journal = {Nature Commun.},
  volume  = {14},
  number  = {1},
  month   = oct,
  year    = {2023},
  pages   = {6675}
}

@ARTICLE{11072420,
author={Enad, Almutasem Bellah and others},
journal={IEEE Commun. Lett}, 
title={Performance Analysis of Data Detection in the {THz}-Band Under Channel-Correlated Noise}, 
year={2025},
volume={29},
number={9},
pages={2098-2102},
doi={10.1109/LCOMM.2025.3586773}}

@ARTICLE{11362947,
  author={Enad, Almutasem Bellah and others},
  journal={IEEE Open J. Commun. Soc.}, 
  title={Diversity Analysis for {Terahertz} Communication Systems Under Small-Scale Fading}, 
  year={2026},
  volume={7},
  number={},
  pages={954-964},
  doi={10.1109/OJCOMS.2026.3656527}}

@mastersthesis{enad2025theoretical,
    author = {ENAD, ALMUTASEM BELLAH AYOUB},
    title = {A Theoretical Performance Analysis Framework for Data Detection in the {Terahertz} Band},
    school = {American University of Beirut},
    year = {2025} 
}

@ARTICLE{11006064,
  author={Dkhan, Ahmad and Tarboush, Simon and Sarieddeen, Hadi and others},
  journal={IEEE Commun. Lett}, 
  title={THz-Band Near-Field {RIS} Channel Modeling for Linear Channel Estimation}, 
  year={2025},
  volume={29},
  number={7},
  pages={1669-1673},
  doi={10.1109/LCOMM.2025.3570839}}

@misc{dkhan2026physicallyconsistentchannelmodeling,
      title={Physically Consistent Channel Modeling and Signal Processing for Reconfigurable Wireless Systems}, 
      author={Ahmad Dkhan and Simon Tarboush and Hadi Sarieddeen and others},
      year={2026},
      eprint={2608.17122},
      archivePrefix={arXiv},
      primaryClass={eess.SP},
}

@misc{enad2026symbolerroranalysislinear,
      title={Symbol Error Analysis of Linear Receivers in {Terahertz} Channels under Channel-Noise Dependence}, 
      author={Almutasem Bellah Enad and others},
      year={2026},
      eprint={2606.15006},
      archivePrefix={arXiv},
      primaryClass={eess.SP},
      url={https://arxiv.org/abs/2606.15006}, 
}

@ARTICLE{enadMImo,
  author={Enad, Almutasem Bellah and Tellambura, Chintha},
  journal={IEEE Commun. Lett}, 
  title={Exact Performance Analysis for Correlated {Rayleigh} {MIMO} Channels}, 
  year={2026},
  volume={},
  number={},
  pages={1-1},
  doi={10.1109/LCOMM.2026.3755072}}

\end{document}